\documentclass[12pt]{article}
\usepackage[utf8]{inputenc}
\usepackage{mathtools}
\usepackage{colortbl} 
\usepackage{lipsum}
\usepackage[dvipsnames]{xcolor}
\usepackage{float}
\usepackage[utf8]{inputenc}
\usepackage{graphicx,setspace,lscape,longtable}
\usepackage{epsfig}
\usepackage{natbib}
\usepackage{amsthm}
\usepackage{bbm}
\usepackage{algorithmic}
\usepackage{algorithm}
\usepackage{authblk}

\usepackage{amssymb, amsmath,bm} 
\numberwithin{equation}{section}
\usepackage{xcolor}         
\usepackage{caption}
\usepackage{subcaption}
\usepackage{hyperref}
\usepackage[english]{babel}
\usepackage{indentfirst}
\usepackage{changebar}
\usepackage{setspace}
\graphicspath{ {images/} } 

\newtheorem{proposition}{Proposition}
\usepackage{soul}
\setstcolor{red}

\usepackage[noabbrev,capitalize]{cleveref}
\crefname{eqnarray}{equation}{equations}
\crefname{equation}{equation}{equations}
\crefname{section}{section}{sections}
\crefname{footnote}{footnote}{footnotes}   
\crefname{line}{line}{lines}   
\crefname{assumption}{assumption}{assumptions}

\numberwithin{equation}{section}

\definecolor{DarkGreen}{rgb}{0.5,0.8,0.6}   
\definecolor{RGBblack}{rgb}{0.0,0.0,0.0}    

\newcommand{\xx}{\color{black} \rm}

\DeclareMathOperator*{\argmax}{arg\,max}

\newcommand{\is}{\itemsep=0pt}
\newcommand{\bdes}[1]{\begin{description}[#1]\is}
  \newcommand{\edes}{\end{description}}
\newcommand{\bi}{\begin{itemize}\is}
  \newcommand{\ei}{\end{itemize}}
\newcommand{\benum}{\begin{enumerate}\is}
  \newcommand{\eenum}{\end{enumerate}}
  \newcommand{\beq}{\begin{eqnarray}\is}
  \newcommand{\eeq}{\end{eqnarray}}

\makeatletter
\newcommand*{\rom}[1]{\expandafter\@slowromancap\romannumeral #1@}
\newcommand{\iidsim}{{\overset{\mathrm{i.i.d.}}{\sim}}}

\newcommand{\bphi}{\bm{\phi}}

\newcommand{\bSigma}{\bm{\Sigma}}

\newcommand{\btheta}{\bm{\theta}}

\newcommand{\bb}{\bm{b}}
\newcommand{\bd}{\bm{d}}
\newcommand{\be}{\bm{e}}
\newcommand{\br}{\bm{r}}

\newcommand{\by}{\bm{y}}

\newcommand{\bz}{\bm{z}}

\newcommand{\bbeta}{\bm{\beta}}

\newcommand{\bx}{\bm{x}}

\begin{document}


\title{Personalized Dynamic Treatment Regimes in Continuous Time: A Bayesian Approach for Optimizing Clinical Decisions with Timing}
\author[1]{William Hua}
\author[2]{Hongyuan Mei}
\author[3]{Sarah Zohar}
\author[4]{Magali Giral}
\author[1, *]{Yanxun Xu}
\affil[1]{Department of Applied Mathematics and Statistics, Johns Hopkins University, Baltimore, USA}
\affil[2]{Department of Computer Science, Johns Hopkins University, Baltimore, USA}
\affil[3]{INSERM, U1138, team 22, Centre de Recherche des Cordeliers, Université Paris 5, Université Paris 6, Paris, France} 
\affil[4]{Centre de Recherche en Transplantation et Immunologie (CRTI), UMR1064, INSERM, Université de Nantes, Nantes, France}
\affil[*]{Correspondence should be addressed to email: yanxun.xu@jhu.edu}
\date{}                     
\setcounter{Maxaffil}{0}
\renewcommand\Affilfont{\itshape\small}
 \doublespacing
  \maketitle



\begin{abstract}
Accurate models of clinical actions and their impacts on disease progression are critical for estimating personalized optimal dynamic treatment regimes (DTRs) in medical/health research, especially in managing chronic conditions. Traditional statistical methods for DTRs usually focus on estimating the optimal treatment or dosage at each given medical intervention, 
but overlook the important question of ``when this intervention should happen.''
We fill this gap by  developing a two-step Bayesian approach to optimize clinical decisions with timing. In the first step, we build a generative model for a sequence of medical interventions---which are discrete events in continuous time---with a marked temporal point process (MTPP) where the mark is the assigned treatment or dosage. Then this clinical action model is embedded into a Bayesian joint framework where the other components model clinical observations including longitudinal medical measurements and time-to-event data conditional on  treatment  histories. In the second step, we propose a policy gradient method to learn the personalized optimal clinical decision that maximizes the patient survival by interacting the MTPP with the model on clinical observations while accounting for uncertainties in clinical observations learned from the posterior inference  of the Bayesian joint model in the first step.  
A signature application of the proposed approach is to schedule follow-up visitations and assign a dosage at each visitation for patients after kidney transplantation.
We evaluate our approach with comparison to alternative methods on both simulated and real-world datasets. 
In our experiments, the personalized decisions made by the proposed method are clinically useful: they are interpretable and successfully help improve patient survival. 
\end{abstract}
\noindent%

    
\section{Introduction}\label{sec:intro}
In biomedical applications involving long-term personalized care of patients with chronic health conditions  (e.g., diabetes, HIV infections, and chronic kidney diseases), 
treatments often include  a sequence of decision making and must be adaptive to the uniquely evolving disease progression of each patient. 
Such scenarios are called dynamic treatment regimes (DTRs). 
Patients with chronic diseases are usually required to follow up with their physicians from time to time and their clinical data are recorded longitudinally.  
Based on these clinical observations, physicians make clinical decisions such as scheduling follow-up visitations and prescribing the right dosages to optimize patient outcomes given a patient's individual characteristics and treatment history at each clinic visitation. Estimating treatment effects and optimizing sequential treatment assignments  from observational data have been extensively studied  in both statistics and machine learning, such as the G-computation formula \citep{robins1986new}, inverse probability of treatment weighting \citep{orellana2010dynamic}, doubly robust methods \citep{zhao2015doubly}, and reinforcement learning \citep{zhao2011reinforcement, clifton2020q}. 
This paper develops a two-step Bayesian approach to optimize clinical decisions with timing. In the first step, 
we develop a Bayesian joint model consisting of a generative probabilistic submodel for clinical decisions with timing and a submodel for clinical observations (e.g., longitudinal clinical measurements and time-to-event data): these two submodels share certain structures and parameters in order to capture the mutual influence between the clinical observations and decisions. The posterior inference of the proposed Bayesian joint model can learn the parameters in the clinical decision submodel as the estimates of how physicians treatment patients in the observed data and uncertainties in clinical observations. 
In the second step, we propose an optimization method that allows the decision model, by interacting with the other parts of the joint model, to learn   to make  the personalized optimal clinical decision at the right time while accounting for uncertainties in clinical observations.   
Such a joint model and the proposed optimization method will be useful in many biomedical applications. 
We elaborate on one signature application in \Cref{sec:sig}, explain why existing methods won't work well on it in \Cref{sec:whynot}, and then give an overview of our method and its technical novelty in \Cref{sec:why}. 


\subsection{A signature application}\label{sec:sig}

A signature medical application of the proposed method would be the kidney transplantation, the most common type of organ transplantation and the primary therapy for  patients with end-stage kidney diseases \citep{Arshad2019}. 
Compared to dialysis, kidney transplantation improves patients' long-term survival and quality of life but with a lower healthcare cost \citep{Jarl2018}. Despite significant advances, a number of complications after surgery still represent important causes of morbidity and mortality for kidney transplant recipients, such as infection, stroke, and graft failure \citep{Lamb2011,Bicalho2019}. To prevent graft rejection, patients are usually hospitalized for a few days initially to monitor signs of complications, then required to have frequent checkups at an outpatient center after being released. At each visitation, they are administered immunosuppressive drugs, such as tacrolimus, to keep their immune systems from attacking and rejecting the new kidney \citep{Kasiske2010}. One crucial medical decision is to schedule  patients' post-transplantation follow-up visitations. 
While follow-up visitation frequency varies from 0-12 months \citep{israni2014variation}, patients with stable kidney function usually have less frequent follow-ups compared to non-stable patients. Another medical decision is to determine the right dosage of tacrolimus at each follow-up visitation since a dosage that is either too high or too low may cause 
serious adverse events. Higher tacrolimus levels have been reported to associate with adverse effects such as neurotoxicity, nephrotoxicity, and cancers \citep{naesens2009calcineurin}; while lower tacrolimus levels are associated with an increased likelihood of graft rejection \citep{Staatz2001}.  
Therefore, optimizing personalized follow-up schedules and prescribing the right dosage of tacrolimus tailored to each patient at each visitation (i.e., precision medicine) are critical and can have a significant impact on patients' survival.

\begin{figure}[ht!]
	\centering
	\includegraphics[page=1,scale=0.45]{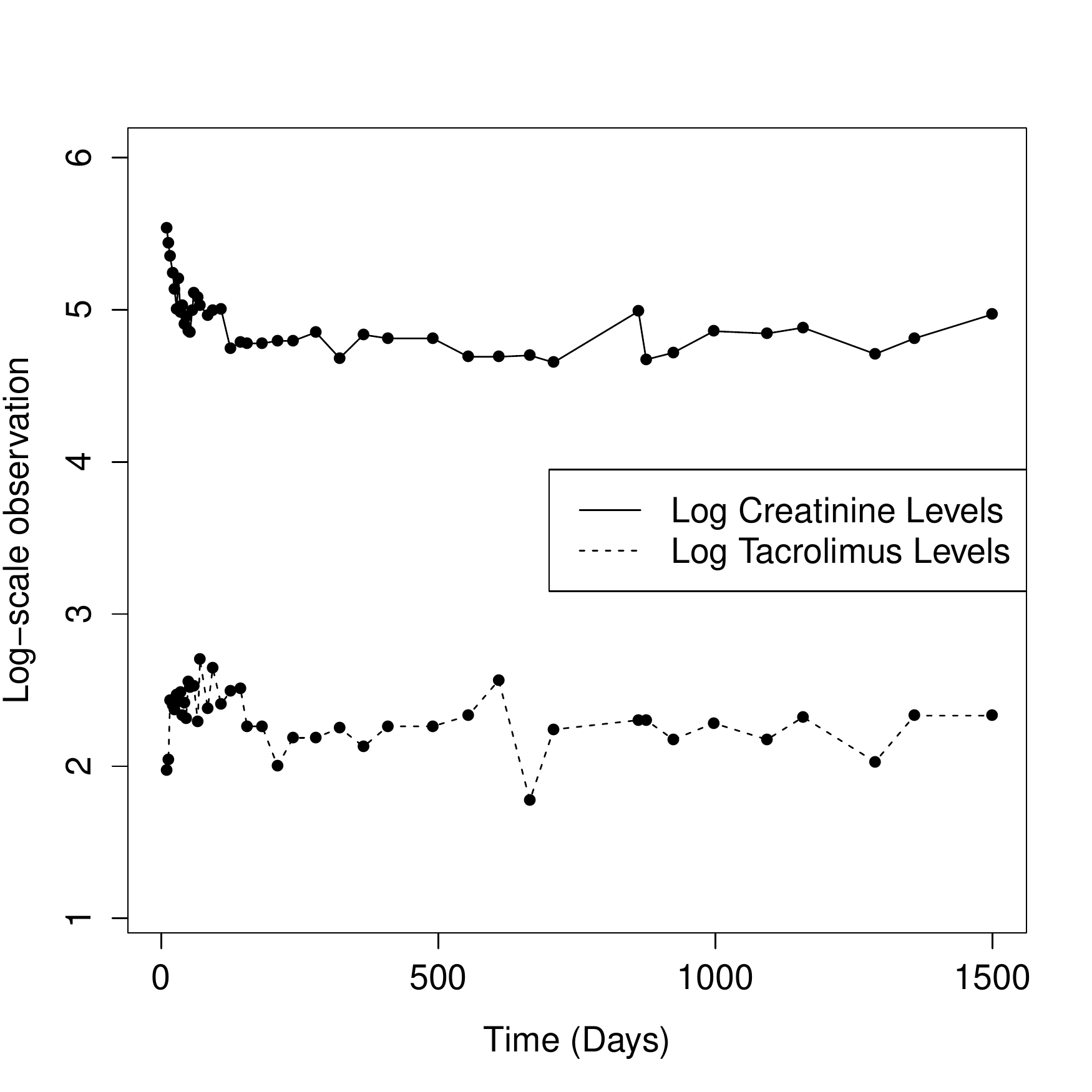}
	\caption{Example data for one patient's creatinine and tacrolimus levels on a log scale over time. The points represent actual visitations. }
	\label{fig:fig1}
\end{figure}

Large-scale kidney transplantation databases, such as French computerized and validated data in transplantation (DIVAT), provide us both opportunities and challenges to determine personalized optimal follow-up schedules and tacrolimus dosages. DIVAT is a database storing
medical records for kidney transplantation in several French hospitals (e.g., Nantes, Paris Necker). Data are collected from the date of transplantation until the graft failure, defined as either returning to dialysis or death with a functioning graft. At each scheduled follow-up visitation, patients' creatinine levels, an important biomarker for measuring kidney function, are collected longitudinally to determine the next follow-up time and assign dosages  by physicians. For example, Figure \ref{fig:fig1} presents one randomly selected patient's longitudinal creatinine levels and tacrolimus dosages versus his/her follow-up visitations  from DIVAT.  In the first several visitations after kidney transplantation, this patient's creatinine levels were high, indicating the kidney was not functioning well; therefore, the physician scheduled a high frequency of follow ups and prescribed high dosages of tacrolimus. As time went by, this patient's kidney function became stable indicated by slowly decreasing creatinine levels; then the prescribed tacrolimus dosages also slowly decreased accompanied with a  decreasing frequency of visitations. 
For patients with kidney transplantation, a major clinical  outcome  of interest is the graft survival time, defined as the time between the transplantation and the first graft failure. 
Follow-up schedules and tacrolimus dosages should be made for the sake of maximizing patients' graft survival time. 

\subsection{Why not use existing methods?}\label{sec:whynot}

Although many statistical and machine learning DTR methods have been developed to optimize sequential clinical decisions \citep{Murphy2003,chakraborty2013statistical,laber2014dynamic, xu2016bayesian,luckett2019estimating}, they don't model, and thus can't optimize, the timing of clinical decisions.  
Most DTR methods regard treatment schedules as known a {\it priori} and only learn to assign optimal treatments at pre-defined schedules. 
For example, \cite{xu2016bayesian} developed a Bayesian nonparametric approach building upon a dependent Dirichlet process and a Gaussian process to determine the optimal treatment regimen  containing a front-line chemotherapy and a salvage treatment   for acute myelogenous leukemia patients.   However, the timing of the salvage treatment was pre-defined as the time when  patients became resistant to the front-line chemotherapy or achieved complete remission first then relapsed.   
\cite{shardell2018joint} proposed a joint mixed effects model using G-computation in which the longitudinal outcome and treatment assignment models share common random effects to estimate causal effects with pre-defined treatment timing. 
\cite{clifton2020q} reviewed the use of Q-learning, a general class of reinforcement learning methods, in estimating optimal treatment regimens   taking the timing of  treatments as given.   \cite{zhao2011reinforcement} attempted to optimize the timing to initiate second-line therapy in the context of clinical trials with two-stage treatments using Q-learning, but only considered two options (i.e., immediately or delayed after induction therapy). Other work on optimizing intervention timing with a fixed number of treatment options include initiation of antiretroviral therapy in HIV \citep{robins2008estimation}, just-in-time adaptive interventions in mobile health\citep{nahum2018just,carpenter2020developments}, and advantage doubly robust policy learning that optimizes when to treat \citep{nie2020learning}.  
\cite{guan2019bayesian} developed a Bayesian nonparametric method that learns to recommend a  regular  recall time for  patients with periodontal diseases. 
However, their method only picks the recall time out of a few pre-defined choices (e.g., 3 months, 6 months, and 9 months) and thus is not applicable to complicated scenarios like the one introduced in \Cref{sec:sig}: at each visitation after kidney transplantation, the next visitation time has to be carefully scheduled given the current clinical measurement in order to maximize the patient's health outcome.  For instance, when patients' kidney function is relatively stable, they should be instructed to wait longer until the next visitation compared to those who are less stable. 
\xx

\subsection{Why use our method?}\label{sec:why}

To the best of our knowledge, the proposed approach is the first general methodology for estimating personalized optimal clinical decisions with timing. 
The method is cutting-edge because (1) we build a generative probabilistic model that properly handles clinical decisions with timing; (2) we embed this decision process into a Bayesian joint model that also models clinical observations; (3)  we propose a Bayesian optimization procedure to optimize personalized treatment schedules alongside other clinical decisions while accounting for uncertainties in clinical observations based on the posterior inference of the proposed Bayesian joint model.    

Our decision model is a marked temporal point process (MTPP) \citep{aalen2008survival}, which is a natural tool to model discrete events in continuous time. 
It has been widely applied and become increasingly popular in various domains, including social science \citep{Butts2017}, medical analytics \citep{Liu2018}, finance \citep{Hawkes2018}, and stochastic optimal control \citep{tabibian2019enhancing}. 
In our example application of \Cref{sec:sig}, each follow-up visitation is an event: the visitation time is assumed to be stochastically scheduled according to the probability distribution characterized by the proposed MTPP;  and the  assigned  tacrolimus dosage, when the visitation happens, is treated as the corresponding  ``mark.'' 

The proposed MTPP for clinical decisions is  then embedded into a Bayesian joint model where it shares certain structures and parameters with the other components modeling clinical observations, including longitudinal creatinine measurements and patient survival in the example application of \Cref{sec:sig}. 
Such design allows our model to capture the complicated mutual influence between clinical observations (e.g., creatinine levels) and decisions (e.g., treatment schedules and tacrolimus dosages). We fit the proposed Bayesian joint model to the data and obtain the posterior inference, 
which not only learns the dynamic patterns of clinical observations with uncertainty quantification, but also estimates the parameters in modeling clinical decisions as the estimates  of  how  physicians  treat patients in practice. The posterior means of these parameters can be used as the initial values in the optimization procedure so that clinical decisions can be refined to  optimize patients' survival.   

\begin{figure}[ht!]
	\centering
	\includegraphics[scale=0.52]{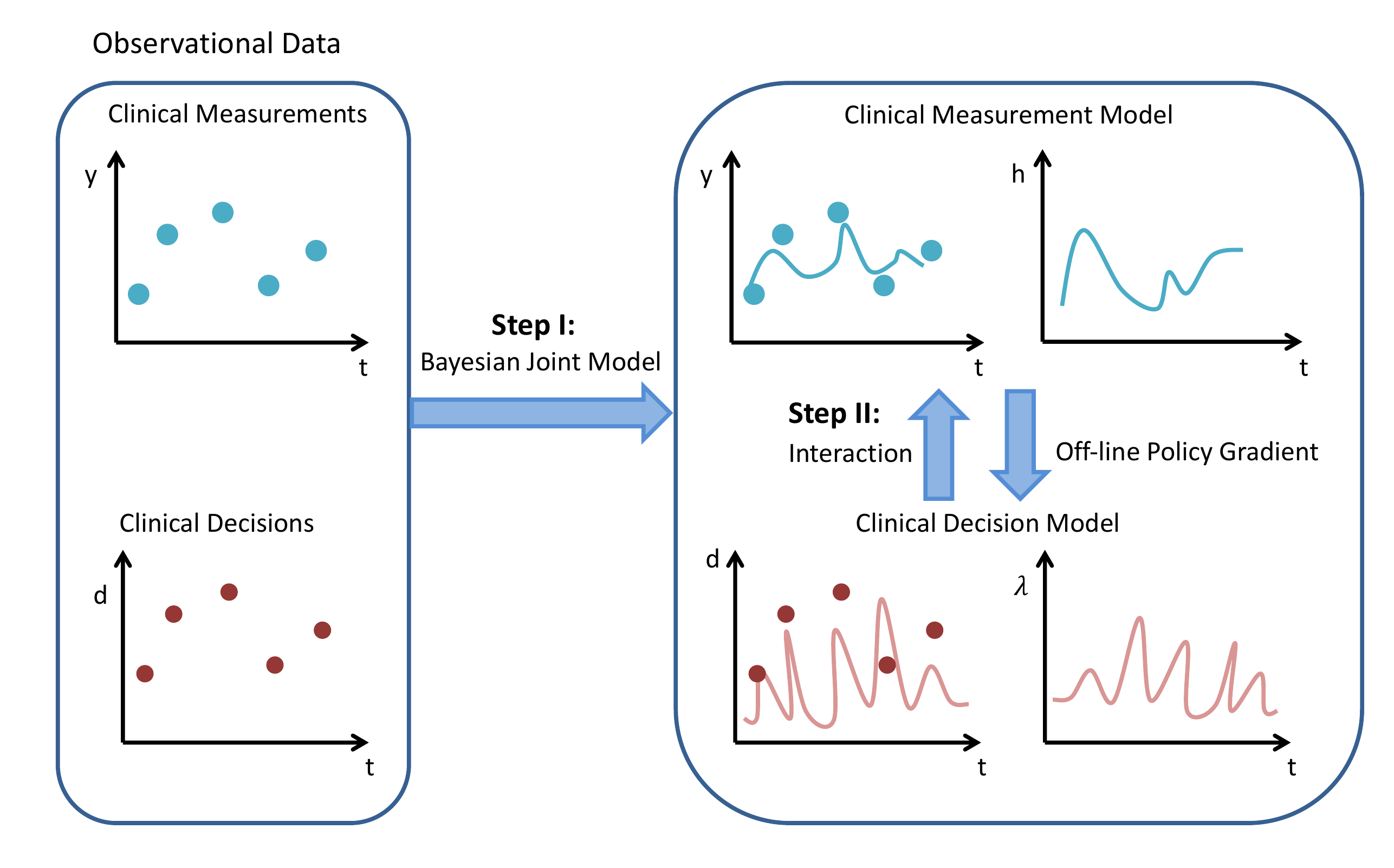}
	\caption{Illustration of the proposed method.  }
	\label{fig:method}
\end{figure}

Next, we let the decision model interact with the observation model in an optimization procedure. 
This technique is known as ``reinforcement learning'' \citep{sutton2018reinforcement}: the decision model (also called the ``policy'') is reinforced, by the feedback from the observation model (also called the ``environment''), to give personalized optimal treatment schedules and dosages that would improve the expected  health outcome for each patient. 
The Bayesian nature of our approach allows the learning  to account for parameter uncertainties in the observation model. 
Figure \ref{fig:method} illustrates the proposed two-step Bayesian approach. 	The left box displays the observational data including the clinical measurements and dosages at a sequence of visitations, and the right-headed arrow next to it stands for the Bayesian joint model (i.e., the first step), pointing to the right box, which represents the learned inference from the Bayesian joint model, including inference on longitudinal measurements, the hazard function of survival, dosages, and the intensity function of visitation timing. 
The vertical arrows inside the right box indicates the policy optimization method (i.e., the second step), in which the decision model learns to achieve higher reward by interacting with the learned measurement model. 
The R package {\it doct} (short for ``Decisions Optimized in Continuous Time")   implementing the proposed model and algorithm is available at 
\url{https://github.com/YanxunXu/doct}. 

The rest of the paper  is organized as follows.   In \Cref{sec:overview}, we outline the overall framework of the proposed two-step Bayesian approach. 
In \Cref{sec:joint}, we elaborate on the first step of developing a Bayesian joint model  
consisting of the decision submodel (for visitation schedules and dosages) and the observation submodel (for clinical longitudinal measurements and patient survival).  
In \Cref{sec:SGD}, we elaborate on the second step of the optimization procedure for the decision model.  
We evaluate  the proposed approach through  simulation studies in \Cref{sec:sim_study} and 
applying it to the DIVAT   kidney transplantation dataset in \Cref{sec:realdata_analysis}. 
Lastly, we conclude the paper with a discussion in \Cref{sec:conclusion}.   

\section{Problem Formulation}
\label{sec:overview}

Our goal is to optimize personalized clinical decision including scheduling a patient' follow-up visitations and prescribing dosages to maximize the patient'  health outcome, e.g., the graft survival time in the kidney transplantation application. 
In this section, we  first introduce the notations, and then outline the framework of the proposed two-step Bayesian approach. 

For each patient $i$, let $T_i$ and $C_i$ denote the graft survival and administration censoring times, respectively, $i=1, \dots, I$. We observe only $\widetilde{T}_i=\min(T_i, C_i)$ and the censoring indicator $\delta_i = \mathbbm{1}(T_i \leq C_i)$.   The graft survival would typically be affected by the patient's baseline risk factors, denoted by $\bx_{i}$, including the patient's age when receiving the transplantation and the donor type.
At each visitation time $t_{i,j}$, the clinical measurement $y_{i,j}$ of interest would be taken, $j=0, \dots, J_i$, and $t_{i,J_i}\leq \widetilde{T}_i$. Note here $t_{i, 0}=0$ denotes the transplantation date of patient $i$, and $y_{i,0}$ denotes the initial creatinine level. Then the physician would prescribe the dosage $d_{i,j}$, and schedule the next visitation time $t_{i,j+1}$. In the kidney transplantation application, $y_{i,j}$ is the logarithm of the creatinine level ($\mu$mol/l), and $d_{i,j}$ is the logarithm of the tacrolimus level (ng/ml). 
We represent the $i$-th patient's sequence of visitations and assigned dosages by $\be_{i, \widetilde{T}_i}=\{(t_{i,0},d_{i,0}), \dots, (t_{i,J_i},d_{i,J_i})\}$. 
Denote $\by_i=(y_{i,0}, \dots, y_{i, J_i})$. Our data can be summarized as $\mathcal{D}=\{\bx_{i}, \widetilde{T}_i, \delta_i, \by_i, \be_{i, \widetilde{T}_i} \}_{i=1}^I$. 

Assume that the physician assigns the dosage $d_{i,j}$ and schedule the next visitation time $t_{i,j+1}$ based on the patient's baseline covariates $\bx_i$ and the longitudinal creatinine measurements $\overline{y_{i,j}}=\{y_{i,j'}: j'\leq j\}$,  we write the joint model of the observed clinical observations and decisions as
\begin{eqnarray}
\label{eq:jointdis}
\prod_{i=1}^I p(\widetilde{T}_i, \delta_i, \by_i, \be_{i, \widetilde{T}_i}\mid \bx_i) &=& \prod_{i=1}^I \Big[p(\widetilde{T}_i, \delta_i \mid \by_i, \be_{i, \widetilde{T}_i}, \bx_i) \nonumber\\
&\times & \prod_j p(y_{i,j}\mid t_{i, j}, d_{i, j-1}, \bx_i)p(d_{i, j}\mid \bx_i, \overline{y_{i, j}}) p(t_{i, j+1}\mid \bx_i, \overline{y_{i, j}})\Big]. \nonumber\\
\end{eqnarray}
We leave the details of the chosen parametrization for this joint model to  \Cref{sec:joint}. Throughput this paper, we use $p(\cdot)$ to denote the probability density. 
In order to identify causal effects, we make the standard assumptions of sequential ignorability and consistency throughout this paper \citep{Murphy2003}. The positivity assumption is satisfied in our model due to the use of the marked temporal point process (MTPP) \citep{aalen2008survival} for modelling stochastic clinical decisions in continuous time, which will be introduced in \Cref{sec:joint1}.

Assume that the joint model \eqref{eq:jointdis} is parameterized with $\btheta$ and $\bphi$, where $\btheta$  denotes the set of parameters related to clinical decisions that control patients' follow-up schedules and dosages at follow-up visitations, and $\bphi$ denotes the remaining parameters. 
We aim to learn the parameters $\btheta$ of the clinical decision model such that, for any (future/hypothetical) patient $i$, a certain reward $R_i$ can be maximized by prescribing the right amount of dosages and scheduling the visitations at the right times. 
Formally, we aim to maximize the expected reward for any patient $i$ with baseline covariates $\bx_i$:  
\begin{eqnarray} 
\label{eq:gi_star21}
G_i(\btheta) &=&  \int E_{ (\by_i, T_i, \be_{i, T_i})\sim p(\by_i, T_i, \be_{i, T_i}\mid \btheta, \bphi, \bx_i) }[R_i]p(\bphi\mid \mathcal{D})d\bphi.
\end{eqnarray} 
Here the expectation is taken over all possible   stochastic   realizations of the clinical measurements $\by_i$, the survival time $T_i$, and the clinical decisions $\be_{i, T_i}$ from their joint distribution $p(\by_i, T_i, \be_{i, T_i}\mid \btheta, \bphi, \bx_i)$;  $\mathcal{D}$ denotes the observed data; and $p(\bphi\mid \mathcal{D})$ is the posterior distribution of $\bphi$ obtained from the posterior inference of the Bayesian joint model \eqref{eq:jointdis}. Note that the reward is stochastic, because the clinical measurements and decisions (i.e., dosages and visitation times), and their influence on the reward are all stochastic. In our kidney transplantation application, we can consider the patient's median survival time as the reward. More details on the reward will be introduced in \Cref{sec:SGD}.

To find the optimal clinical decision parameter $\widetilde{\btheta}_i$ that maximizes the expected reward $G_i(\btheta)$ for patient $i$ after integrating out the uncertainty in the longitudinal process and the survival distribution,
we propose a two-step Bayesian approach. In the first step that will be introduced in \Cref{sec:joint}, we will fit the joint model \eqref{eq:jointdis} to the observed data and assign priors for parameters $\btheta$ and $\bphi$, then obtain posterior inference through Markov chain Monte Carlo (MCMC) simulations. 
In the second step that will be introduced in  \Cref{sec:SGD}, we will propose a policy gradient method using stochastic gradient descent (SGD) \citep{Ruder2016} to optimize personalized clinical decision. The uncertainties in the clinical observations will be incorporated by integrating over the posterior distribution of $p(\bphi\mid \mathcal{D})$ when calculating the reward \eqref{eq:gi_star21}. The estimated posterior mean of $\btheta$ from the first step can be used as an initial value in the optimization procedure for efficient learning. 

\section{First Step: A Bayesian Joint Model}
\label{sec:joint}


In this section, we describe the proposed Bayesian joint model for both clinical decisions and observations. 
In \Cref{sec:joint1}, we introduce the clinical decision model  for follow-up visitation  schedules and dosages; in \Cref{sec:joint2}, we introduce the clinical  observation model for longitudinal measurements and time-to-event data, which are linked to the   decision model through parameter sharing. 
To facilitate our presentation and readers' understanding, we will use the kidney transplantation example and the DIVAT data to illustrate the model. 
However, the proposed method is applicable to general medical settings since the patterns that the method can capture are not tied to this particular application.

\subsection{Modeling clinical decisions}
\label{sec:joint1}
Modeling event data with marker information is important to learn the latent mechanisms that govern the observed stochastic event patterns over time in many domains, such as social science \citep{Butts2017} and medical analytics \citep{Liu2018}. 
Marked temporal point processes  \citep{aalen2008survival} are a general framework for modeling such event data. Formally, a marked temporal point process is a random process, the realization of which consists of a sequence of events localized in time, i.e., $\mathcal{H}=\{(t_0, d_0), (t_1, d_1), \dots, (t_J, d_J)  \}$ with the occurrence time of event $j$ being $t_j\in \mathbb{R}^+$ and $d_j$ is the associated mark. In our application, $t_j$ represents the time when a patient visits an outpatient center and $d_j$ represents the tacrolimus dosage assigned by the physician. The first event is defined as the day of transplantation at $t_0=0$ with an initial dose $d_0$. 

\begin{figure}[ht!]
	\centering
	\begin{tabular}{cc}
		\includegraphics[page=1,width=.46\textwidth]{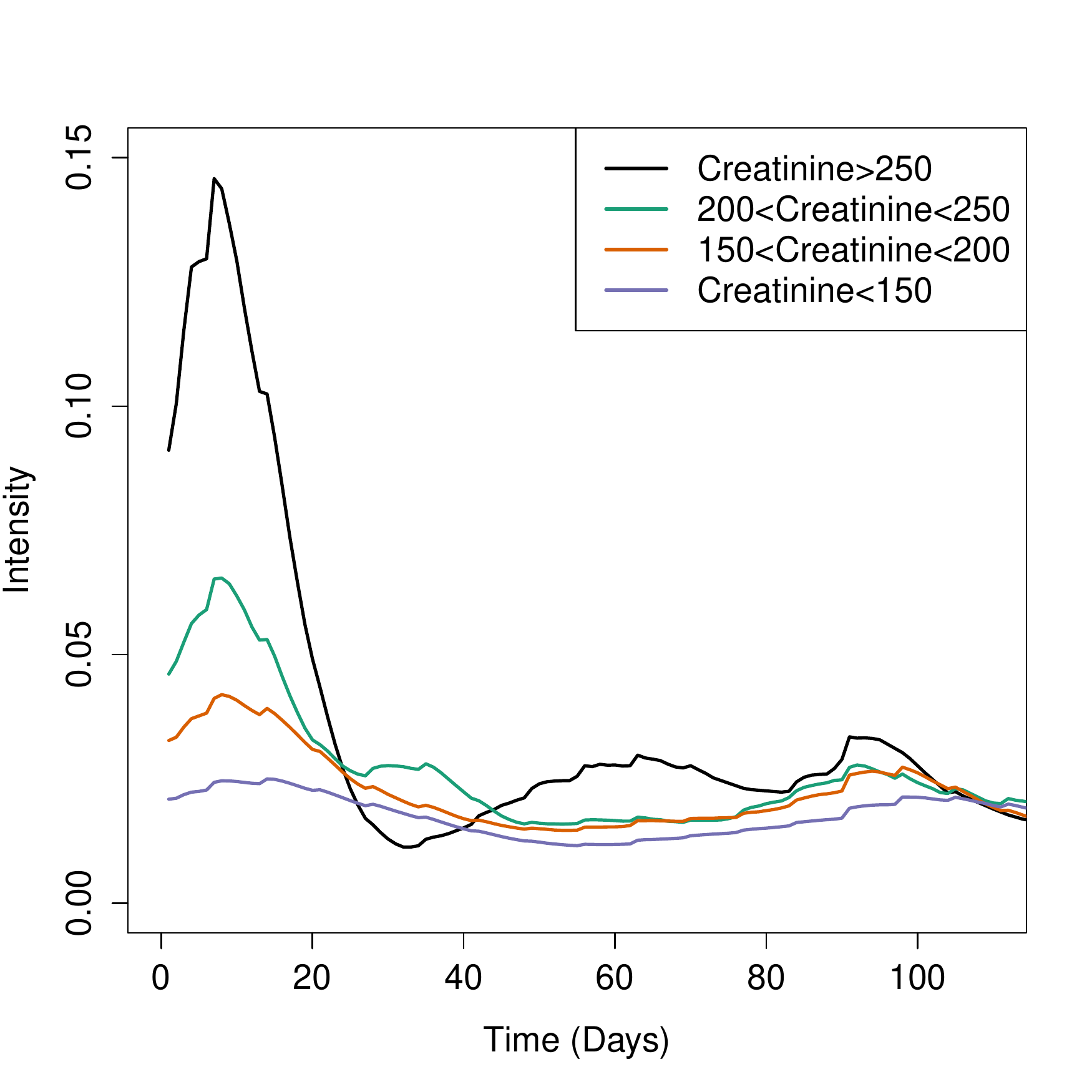} &
		\includegraphics[page=1,width=.48\textwidth]{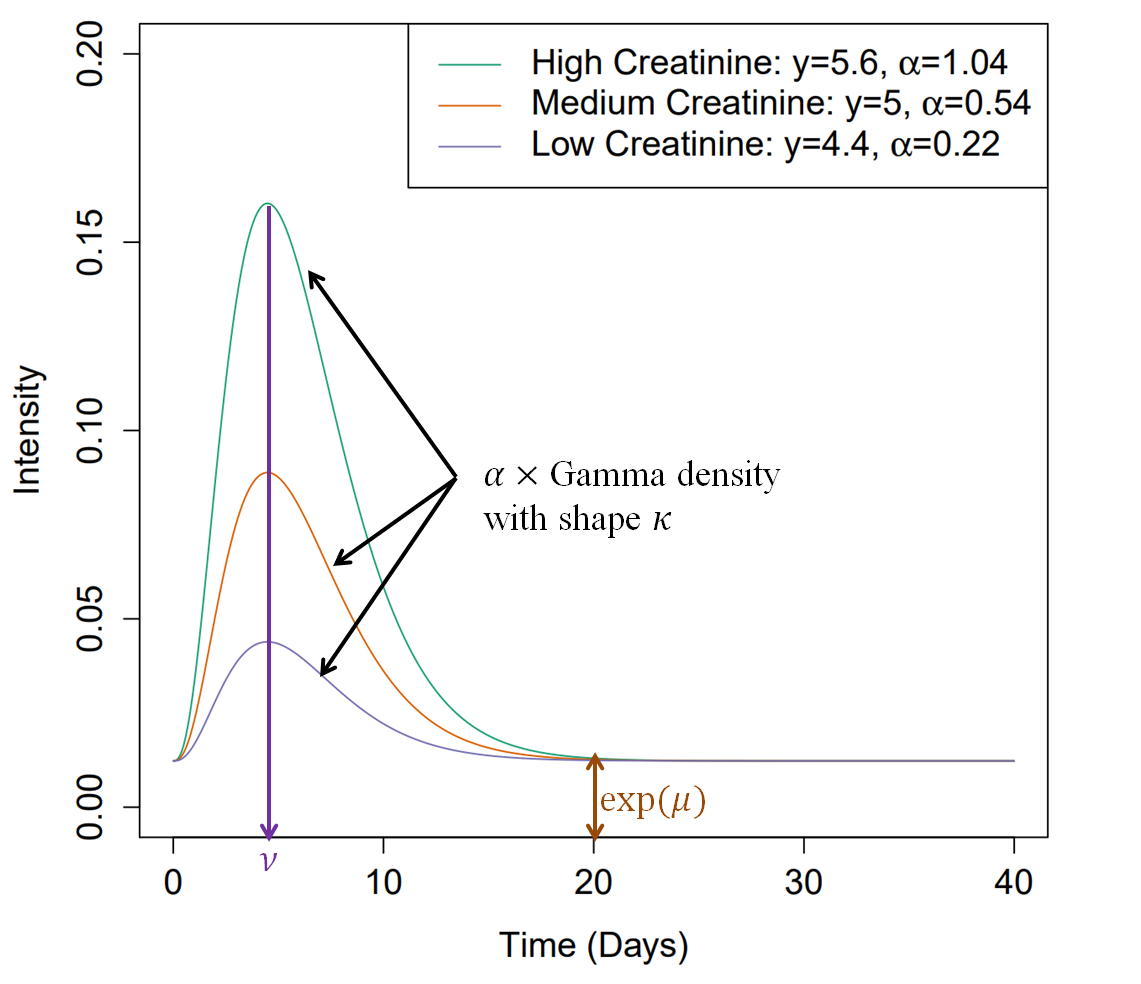}\\
		(a) Empirical intensity & (b) The proposed intensity\\
	\end{tabular}
	\caption{Panel (a) shows the empirical intensity plot for the amount of time (in days) between follow-up visitations. Panel (b) plots an example of how creatinine levels and model parameters affect the visitation intensity, where  $\xi=2$, $\bbeta_{\alpha}=(10,-1.8)^T, \mu=-4.4 $, $\nu_1=1.5$, and $\nu_2=1$.}
	\label{fig:fig3} 
\end{figure}

Denote the event history up to time $t$ to be $\mathcal{H}_t=\{(t_j, d_j)\in \mathcal{H}\mid t_j<t\}$. Under MTPP, the instantaneous rate of the event is characterized by a  conditional intensity function $\lambda(t)$, namely $\lambda(t) = \lim_{dt\rightarrow 0} \frac{Pr \{ \mathrm{event\ happens \ in} \ [t, t+dt)\mid \mathcal{H}_t\}}{dt}$. 
Common forms of the conditional intensity function $\lambda(t)$ include Poisson process \citep{Zhu2018}, Gamma process \citep{Shibue2020}, Hawkes process \citep{hawkes-71}. 
However, these common models cannot capture complicated patterns in many medical applications. For instance, 
as shown in \cref{fig:fig3}(a) that plots the empirical 
intensity of the amount of time between visitations for different ranges of creatinine levels in the DIVAT data, the elapsed time between follow-up visitations depends on the creatinine level. Also, the empirical intensities of visitations are observed to quickly rise to a peak and then fall down accompanied by moderate oscillations. 
Such complication is beyond the capacity of the Poisson process that assumes a constant intensity and the Gamma process whose intensity function is monotonic. 
The Hawkes process assumes that the past events always elevate the intensities of future events and this ``self-exciting'' effect is additive---it is also apparently not the dynamics that the visitations in the DIVAT data actually follow. 

We propose a flexible conditional intensity function that  also incorporates human intuition: it   takes  longitudinal clinical measurements into account and captures patients' heterogeneity. 
Recall that $y_{i,j}$ denotes 
the logarithm of the creatinine level for patient $i$ at the $j$-th follow-up visitation occurring at time $t_{i,j}$ (days). 
Our conditional intensity function makes use of a Gamma density function as follows: 
\begin{equation}
\lambda_{i}(t)= \underbrace{\exp(\mu)}_\text{Baseline Intensity} + \ \ \alpha_{i,j} \underbrace{(t-t_{i,j})^{\kappa-1} e^{-\gamma(t-t_{i,j})} \frac{\gamma^{\kappa}}{\Gamma(\kappa)} }_\text{Gamma density} \text{ for } t \in (t_{i,j} , t_{i, j+1}], \label{eqn:intensity}
\end{equation}
where $\alpha_{i,j}>0,\gamma>0, \kappa\geq 1$. 
The parameter $\alpha_{i,j}$ is patient-specific so that our intensity function $\lambda_{i}$ is personalized. 
We set $\kappa=\exp(\nu_2)+1 > 1$ so that the intensity   rises to  a ``global peak''   
and then decreases: it would eventually approach to the ``baseline level'' $\exp(\mu)$ unless the next visitation happens and sets up a new intensity curve.
For easy interpretation, we parameterize $\gamma$ as $\gamma = \exp(\nu_2 - \nu_1)$ such that the ``peak time''  (i.e., when the peak of the intensity function occurs)
can be easily computed as $\frac{\kappa-1}{\gamma}=\exp(\nu_1)$. 
Moreover, since the intensity level  often depends on the clinical 
measurement (e.g., as in \cref{fig:fig3}(a), a higher creatinine level implies a  higher intensity), we condition the parameter $\alpha_{i,j}$, which controls the peak intensity for patient $i$ between time $t_{i, j}$ and $t_{i, j+1}$, on the clinical measurement taken at the $j$-th visitation:  
$$\alpha_{i,j}= \frac{\xi}{1+\exp((1,y_{i,j}) \bbeta_\alpha)}.$$
This design reflects the human intuition that the time of ``next visit'' is usually determined based on the clinical measurement of ``this visit.'' 
Note that our design allows incorporating other covariates (i.e., measurements) by simply augmenting them to the vector $(1, y_{i,j})$. 
Figure \ref{fig:fig3}(b) shows how the visitation intensity under our model is affected by the most recent creatinine level $y_{i,j}$ (and thus the magnitude parameter $\alpha_{i,j}$) given a specific set of parameter values.

Next, we model the dosage $d_{i,j}$  at the $j$-th visitation of patient $i$ as the ``mark'' of the visitation event. 
Generally speaking, the physician would assign a dosage based on the patient's current clinical measurement $y_{i,j}$ and potential risk factors $\bx_{i}$. We assume the following dosage model reflecting this knowledge:
\begin{equation}
d_{i,j}=  (1,y_{i,j}, \bx_{i} ) \bbeta_{d}  + \zeta_{i,j},
\label{eqn:dosage}
\end{equation}
where $\zeta_{i,j} \iidsim \mathrm{Normal}(0,\sigma_{d}^2)$. This Gaussian error assumption is suitable for our dosage data, which has continuous values, but can be easily modified for other dosage/treatment types (e.g., generalized linear regression models for discrete treatment choices). 
In the kidney transplantation application,  
we assume that the clinical decision at time $t_{i,j}$ (i.e., assigning a dosage $d_{i,j}$ and scheduling the next visitation time $t_{i, j+1}$) is independent of the patient's history conditional on the measured creatinine at the current visit $y_{i,j}$. This is reasonable in our application since the duration between two consecutive visitations for patients after kidney transplantation can be months or even longer. However, this assumption can be relaxed as in other popular DTR methods such as Q-learning \citep{clifton2020q} if desired. For instance,  the patient's past creatinine levels can be included in the vector $(1,y_{i,j}, \bx_{i} )$ of \eqref{eqn:dosage} to model the effect of past creatinine levels on the dosage. 
Thus, the probability density of the $i$-th patient's sequence of visitations and assigned dosages $\be_{i, T_i}=\{(t_{i,0},d_{i,0}), \dots, (t_{i,J_i},d_{i,J_i})\}$ up to 
time $T_i$ can be written as 
\begin{equation} 
\label{eq:dos_vis_like}
\begin{gathered}
p(\be_{i,T_i}\mid \by_i,\bx_i, \bbeta_v,\bbeta_d,\sigma_d^2)\\
=\underbrace{\exp\Big(-\int_0^{T_i}\lambda_{i}(t \mid \by_i, \bbeta_v )dt\Big)}_{\text{Prob. of no visits at } t \in [0,T_i] \backslash \{t_{i,j}\}_{j=1}^{J_i} } 
\prod_{j=0}^{J_i} 
\underbrace{p(d_{i,j}\mid y_{i,j}, \bx_i, \bbeta_d,\sigma_d^2)}_{\text{Prob. of dosage }  } 
\prod_{j=1}^{J_i} 
\underbrace{ \overbrace{\lambda_{i}(t_{i,j}  \mid y_{i,j-1}, \bbeta_v )}^{ \eqref{eqn:intensity} } }_{\text{Prob. of a visit at } t_{i,j}}   ,
\end{gathered}
\end{equation}
where $\by_i=(y_{i,0}, \dots, y_{i, J_i})$, $\bbeta_v= \{\mu,\nu_1,\nu_2, \xi, \bbeta_\alpha \}$. 

\subsection{Linking clinical observations with clinical decisions}
\label{sec:joint2}

In this section, we introduce the proposed Bayesian joint model that links the submodel of the  longitudinal measurements and time-to-event data to the MTPP  of the clinical decisions by carefully designing parameter sharing in order to capture the mutual influence between clinical observations and decisions. 
Shortly in \Cref{sec:SGD}, we will leverage this joint model to optimize clinical decisions with the goal of maximizing patients' survival.



Our clinical observation model is composed of two submodels---a linear mixed effects model for longitudinal clinical measurements (e.g., creatinine levels) and a time-to-event model for patient survival (e.g., graft survival time after kidney transplantation). 
The two submodels are then connected by sharing random effects \citep{Rizopoulos2014}. 
Recall that $y_{i,j}=y_i(t_{i,j})$ denotes the longitudinal measurement value   for patient $i$ at $j$-th follow-up visitation at time $t_{i,j}$, $i=1, \dots, I$,  $j=0, \dots, J_i$. 
Let $y_i^*(t)$ be the underlying true but unobserved longitudinal process at time $t\geq 0$. We assume
\begin{equation}\label{eqn:true_creatinine}
y_i(t) = y_i^*(t) + \epsilon_{i,j} = \bz_i(t) \bbeta_{l}   + \br_i(t) \bb_i+ \epsilon_i(t), 
\end{equation}
where $\epsilon_{i}(t) \iidsim \mathrm{Normal}(0,\sigma_{l}^2)$ and  $\bb_i \sim \mathrm{Normal}({\bf 0}, \bSigma_b)$. The covariate vectors $\bz_i(t)$ and $\br_i(t)$ are associated with fixed and random effects respectively: 
\begin{equation*}\label{eqn:effects}
\bz_i(t)=( 1,d_i(t), \bx_{i}, t,t^2) \text{ and } \br_i(t)=(1,d_i(t),t),
\end{equation*}
where $d_i(t)$ at time $t$ is the dosage assigned by the physician at the most recent visitation, i.e., 
$d_i(t) = d_{i,j} \text{ for } t \in (t_{i, j}, t_{i, j+1}].$
The temporal dependence of $\bz$ and $\br$ on the dosage $d$ captures the   drug effect on the longitudinal  measurements   of interest: in the kidney transplantation application, it is supposed to capture the suppressive effect of tacrolimus on the creatinine level.  
Denote $\bd_i=(d_{i,0}, \dots, d_{i, J_i})$,  the probability of the observed sequence of creatinine measurements $\by_i$ is 
\begin{equation} 
\label{eq:long_like}
\begin{gathered}
p(\by_i\mid \bd_{i},\bx_i,   \bbeta_l,\sigma_l^2,\bb_i)  = \prod_{j=1}^{J_i} p(y_{i,j} \mid t_{i,j}, d_{i, j-1}, \bx_i,  \bbeta_l,\sigma_l^2,\bb_i).
\end{gathered}
\end{equation}

Next, we construct the time-to-event submodel depending on the underlying true longitudinal trajectory $y^*_{i}(t)$ and the MTPP that models clinical decisions. We consider a Weibull proportional hazards model  as follows:
\begin{equation} 
\label{eq:surv_haz}
\begin{gathered}
h_{i}(t)=\exp\Big(-(\underbrace{\beta_{s1} y_i^*(t)}_\text{longitudinal effect} + \underbrace{ \beta_{s2} d_{i}(t) + \beta_{s3} \mathrm{Tox}_i(t) }_\text{dosage effect} + \underbrace{\beta_{s4} \alpha_{i}(t)}_\text{visitation effect} +h_0 )\Big)\omega t^{\omega-1}, \end{gathered}
\end{equation}
where  $\omega$  is the shape parameter. If desired,  more complex survival models can be  explored, such as Cox proportional hazard models \citep{lin1989robust} and Bayesian nonparametric survival regression models \citep{xu2019bayesian}. 
The dependence on $y_i^*(t)$ reflects the domain knowledge that the survival event 
is usually associated with the underlying health condition reflected by longitudinal measurements. 
The dosage effect term in \cref{eq:surv_haz} measures the overall   drug   effect  on the patient: 
$\beta_{s2}d_i(t)$ is the ``instantaneous'' effect while $\beta_{s3} \mathrm{Tox}_i(t)$ is the ``accumulated'' effect: 
\begin{equation*}
\mathrm{Tox}_i(t)=\int_0^{t} d_i(\tau)\eta_{tox} \exp(-(t-\tau)/\eta_{tox}) d\tau,
\end{equation*}
where the parameter $\eta_{tox}$ controls the rate of the exponential weighting for the past dosages. 
In practice, the instantaneous effect is usually beneficial (e.g., tacrolimus reduces the likelihood of graft rejection or death) while the accumulated effect is often toxic (and that is why we name it $\mathrm{Tox}$): 
e.g., a prolonged high dosage of tacrolimus might have adverse effects on kidneys, central nervous system, and gastrointestinal tract, thereby worsening a patient's survival \citep{randhawa1997}.
We also link the survival submodel with the visitation model by defining $\alpha_{i}(t) = \alpha_{i,j} \text{ for } t \in (t_{i, j}, t_{i,j+1}]$ since a high visitation intensity (i.e., larger $\alpha_{i,j}$) typically implies a higher risk, e.g., graft failure and thus shorter expected survival time.

Recall that $T_i$ and $C_i$ denote the survival and censoring times for patient $i$, respectively. We assume $\widetilde{T}_i=\min(T_i, C_i)$ and $\delta_i = \mathbbm{1}(T_i \leq C_i)$.  Denote $f_i(t)$ and $S_i(t)$ to be the corresponding density and survival functions of the hazard function \eqref{eq:surv_haz}: $S_i(t)=\exp(-\int_0^{t} h_i(u)du)$, $f_i(t)=h_i(t)S_i(t)$.  We can write the survival likelihood for patient $i$ as
\begin{eqnarray} 
\label{eq:surv_like}
p(\widetilde{T}_i,\delta_i\mid \by_i, \bx_i, \be_{i, \widetilde{T}_i} , \bbeta_l,\bb_i,\bbeta_s) 
&=&f_i(\widetilde{T}_i\mid \by_i,\bx_i,  \be_{i, \widetilde{T}_i},\bbeta_l,\bb_i,\bbeta_s)^{\delta_i} \nonumber\\
&&\times S_i(\widetilde{T}_i\mid \by_i, \bx_i, \be_{i, \widetilde{T}_i},\bbeta_l,\bb_i,\bbeta_s)^{1-\delta_i} ,
\end{eqnarray}
where $\bbeta_s= \{\omega, \beta_{s1}, \beta_{s2} , \beta_{s3} , \beta_{s4},  h_0, \eta_{tox},\bbeta_\alpha, \xi\}$. 
In summary,  we propose a joint model consisting of an MTPP for clinical decisions including 
follow-up visitation schedules   and dosages, 
a linear mixed effects model for longitudinal clinical measurements, 
and a time-to-event model for the patient survival; they are inter-connected by sharing structures and parameters. 
The joint probability of the clinical observations and decisions can then factor as
{  
	\begin{align}   
	& \prod_{i=1}^{I}  p(\by_i,\be_{i, \widetilde{T}_i}, \widetilde{T}_i,\delta_i \mid \bx_i, \bbeta_l,\bbeta_d,\bbeta_v,  \bbeta_s,\bb_i,\sigma_l^2,\sigma_d^2) \nonumber\\
	\propto& 
	\prod_{i=1}^{I} \Bigg(
	\underbrace{ p(\be_{i, \widetilde{T}_i}\mid \by_i,\bx_i, \bbeta_v,\bbeta_d,\sigma_d^2) }_\text{ \eqref{eq:dos_vis_like}} 
	\underbrace{p(\by_i\mid \bd_i, \bx_i,  \bbeta_l,\sigma_l^2,\bb_i) }_\text{ \eqref{eq:long_like}} 
	\underbrace{p(\widetilde{T}_i,\delta_i\mid  \by_i,\bx_i,  \be_{i, \widetilde{T}_i},\bbeta_l,\bb_i,\bbeta_s)  }_\text{\eqref{eq:surv_like}}
	\Bigg). \label{eq:like_all}
	\end{align}   
}
Note that \eqref{eq:like_all} appear to have circulation of $\bd_i \mid \by_i$ in  \eqref{eq:dos_vis_like} and $\by_i \mid \bd_i$ in \eqref{eq:long_like}. However, it is a valid factorization because the dependencies are temporal: each dosage $d_{i,j}$ is conditional on  the \emph{current} measurement $y_{i,j}$ in 
$\bd_i \mid \by_i$, while the \emph{next} measurement $y_{i, j+1}$ is conditional on the current dosage $d_{i,j}$ in $\by_i \mid \bd_i$. 



We complete the model by imposing the following priors:  $\bbeta_{d} \sim \mathrm{Normal}(\bbeta_{d0}, \bSigma_{\bbeta_{d}}), $ $  \sigma_{d}^2 \sim \mathrm{InverseGamma}(\pi_{d1},\pi_{d2})$, $ \bbeta_{l} \sim  \mathrm{Normal}(\bbeta_{l0},\bSigma_{\bbeta_{l}}), \ \sigma_{l}^2 \sim \mathrm{InverseGamma}(\pi_{l1},\pi_{l2})$ for conjugacy. We assume a flat prior for  $\bSigma_{b}$. When conjugacy is unattainable for the visitation and survival parameters, we assume $\beta_{s1}, \beta_{s2} , \beta_{s3} , \beta_{s4}, h_0 \sim \mathrm{Normal}(\beta_{s0}, \sigma_{s0}^2) $, 
$ \eta_{tox}\sim \mathrm{Gamma}(\pi_{s1},\pi_{s2})$, $ \omega \sim \mathrm{Gamma}(\pi_{s3},\pi_{s4})$, $\mu, \nu_{1}, \nu_{2} \sim \mathrm{Normal}(\beta_{v0}, \sigma_{v0}^2)$, $\bbeta_{\alpha} \sim \mathrm{Normal}(\bbeta_{\alpha0}, \bSigma_{\bbeta_{\alpha} }), $ and $\xi \sim \mathrm{Gamma} (\pi_{v1},\pi_{v2})$. We carry out posterior inference using the 
Markov chain Monte Carlo (MCMC) sampler. The details are included in the Supplementary Material Section A.

\section{Second Step: Optimize Personalized Clinical Decision}
\label{sec:SGD}
In this section, we  
propose a policy gradient method using stochastic gradient descent (SGD) \citep{Ruder2016} to optimize personalized clinical decision including scheduling a patient' follow-up visitations and prescribing dosages to maximize the patient'  health outcome, e.g., the graft survival time in the kidney transplantation application.

Let $\btheta=(\nu_{1},\nu_{2},\mu, \bbeta_d, \sigma_d^2 )$  denote the set of ``policy'' parameters related to clinical decisions, i.e., the parameters that only appear in the 
conditional intensity function \eqref{eqn:intensity} and the mark distribution \eqref{eqn:dosage}, which control patients' follow-up schedules and dosages at follow-up visitations. 
Let $\bphi=(\bbeta_{s},\bb_i,\bbeta_l, \sigma_l^2 )$   denote the set of ``observation'' parameters, i.e.,  all other parameters in the joint model \eqref{eq:like_all}. 
Recall that in \cref{sec:overview}, we define the expected reward for any future/hypothetical  patient $i$ with baseline covariates $\bx_i$ to be: 
\begin{eqnarray} 
\label{eq:gi_star2}
G_i(\btheta) &=&  \int E_{ (\by_i, T_i, \be_{i, T_i})\sim p(\by_i, T_i, \be_{i, T_i}\mid \btheta, \bphi, \bx_i) }[R_i]p(\bphi\mid \mathcal{D})d\bphi. 
\end{eqnarray} 
The expectation is taken over all possible   stochastic   realizations of $(\by_i, T_i, \be_{i, T_i})$. 
In the kidney transplantation application, we define a personalized reward function $R_i$ as the log-scaled median survival time to optimize patients' survival: $R_i=\log(\widehat{T}_i)$, where $S_i(\widehat{T}_i)=0.5$. This reward provides computational and variance-reduction advantages over a naive choice of the survival time itself. If desired, other reward functions can be considered. For example, if a physician or patient would like to take into consideration the healthcare cost per visit, we could penalize the number of visitations in the reward function, e.g., $R_i=\log(\widehat{T}_i)+\eta_0 C_i$, where $\eta_0$ is a tuning parameter and $C_i$ is the number of visitations. 
Our goal is to find, for any patient $i$ with baseline covariates $\bx_i$, 
the optimal clinical decision, represented by $\widetilde{\btheta}_i$, to maximize the expected reward $G_i(\btheta)$ after integrating out the uncertainty in the longitudinal process and the survival distribution: 
$$\mathrm{maximize}_{p(\be_{i, T_i}\mid \btheta)}  G_i(\btheta), $$
where $p(\be_{i, T_i}\mid \btheta)$ is the probability density of the MTPP. 

To  find the optimal clinical decision parameter $\widetilde{\btheta}_i$ for patient $i$, we use stochastic gradient descent (SGD) \citep{Robbins1951}, i.e., $\btheta_{i,m+1}=\btheta_{i,m}+s_{i,m}\nabla_{\btheta}G_i(\btheta)\mid_{\btheta=\btheta_{i, m}}$, which requires computing the gradient of the expected reward: $\nabla_{\btheta}G_i(\btheta)$.  As the expectation is taken over realizations of the joint distribution $p(\by_i, T_i, \be_{i, T_i}\mid \btheta, \bphi, \bx_i)$, it is intractable to directly  compute $\nabla_{\btheta}G_i(\btheta)$. 
Fortunately, we can indirectly compute this gradient by taking the expectation of the reward-weighted gradient of log-policy. Precisely, 
\begin{proposition} 
	For any patient i with baseline covariates $\bx_i$, given  a joint distribution $p(\by_i, T_i, \be_{i, T_i}\mid \btheta, \bphi, \bx_i)$, the gradient of the expected reward $G_i(\btheta)$ with respect to $\btheta$ is: 
	$$\nabla_{\btheta} G_i(\btheta)=  \int  E_{ (\by_i, T_i, \be_{i, T_i})\sim p(\by_i, T_i, \be_{i, T_i}\mid \btheta, \bphi, \bx_i) } [R_i \nabla_{\btheta} \log p(\be_{i, T_i} \mid \by_i,\bx_i,\bphi,\btheta) ] p(\bphi\mid \mathcal{D})d\bphi, $$
	\label{thm:gi_star_grad}
	where $p(\be_{i, T_i} \mid \by_i,\bx_i,\bphi,\btheta)$ is the probability of the patient’s sequence of visitations and assigned dosages in  \eqref{eq:dos_vis_like}.
\end{proposition}
We leave the detailed proof of \Cref{thm:gi_star_grad} to Supplementary Section B.

According to \Cref{thm:gi_star_grad}, in order to compute $\nabla_{\btheta} G_i(\btheta)$, we first need to be able to sample  $\by_i, T_i, \be_{i, T_i} $ from $p(\by_i, T_i, \be_{i, T_i}\mid \btheta, \bphi)$  and calculate $R_i$ from the generated samples. We sample the $j$-th follow-up visitation time $t_{i,j}$ and the survival time $T_i$ using an inverse transform sampling method: first computing the cumulative distribution function (CDF) of the distribution, sampling a random number $U$ from $\mathrm{Uniform}(0, 1)$, and then 
inverting the CDF function at $U$ to yield the visitation/survival time \citep{Giesecke2011}. If the $j$-th visitation time occurs before the survival time, i.e., $t_{i,j}<T_i$, we sample $y_{i,j}$ and $d_{i,j}$ from their respective distributions and continue to sample the $(j+1)$-th  visitation time and the  survival time. We iteratively sample follow-up visitation times, survival times, longitudinal measurements, and dosages until the sampled survival event occurs before the next visitation time. After obtaining samples of $\by_i, T_i, \be_{i, T_i}$, we can easily compute $R_i$. 
We describe the sampling process for a general $R_i$  in Algorithm \ref{sample_alg}. The algorithm details of sampling $\by_i, T_i, \be_{i, T_i}, R_i$   for the reward being the log median survival time are provided in   Supplementary Section C. 

\begin{algorithm}[ht!]
	\caption{Sampling $\by_i, T_i, \be_{i, T_i}$  from the joint model and  computing $ R_i$ }
	\label{sample_alg}
	Let $\be_{i,T_i}=\{(t_{i,0},d_{i,0}), \dots, (t_{i,J_i},d_{i,J_i})\}$, and $\by_i= (y_{i}(t_{i,1}),\dots,y_{i}(t_{i,J_i}))$ denote the simulated follow-up schedules, dosages, and longitudinal data over $J_i$ visitations  until the  survival time, $T_i$ for any patient $n$ with covariates $\bx_i$. \\
  \textbf{Input}: $\btheta$, $\bphi$, $\bx_i$, $y_{i,0}$ \\
  \textbf{Output}: $\by_i$, $T_i$, $\be_{i,T_i}$, $R_i$ 
	\begin{algorithmic}[1]
		\STATE Initialize $j \gets 1$, continue $\gets \textbf{true} $
		\STATE $t_{i,0} \gets 0$
		\STATE $y_{i}(0) \gets y_{i,0}$
		\STATE $d_{i,0} \gets \mathrm{Normal}( (1,y_{i}(0), \bx_{i} ) \bbeta_d, \sigma_d^2)$ 
		\STATE $ \bb_i \gets  \mathrm{Normal}({\bf 0}, \bSigma_b)$
		\WHILE{continue}
		\STATE $U_v \gets \mathrm{Uniform}(0,1)$
		\STATE Solve for $t_{i,j}$ : $1-\exp(-\int_{t_{i,j-1}}^{t_{i,j}} \lambda_i(x)dx)=U_v$
		\STATE $U_s \gets \mathrm{Uniform}(0,1)$
		\STATE Solve for $T_i$ : $1-\exp(-\int_{t_{i,j-1}}^{T_i} h_i(x)dx)=U_s$
		\IF{$T_i>t_{i,j}$} 
		\STATE $\bz_i(t_{i,j}) \gets ( 1,d_i(t_{i,j-1}), \bx_{i}, t_{i,j},{t_{i,j}}^2)$,  $\br_i(t_{i,j}) \gets (1,d_i(t_{i,j-1}),t_{i,j})$
		\STATE $ y_{i}(t_{i,j}) \gets \mathrm{Normal}(\bz_i(t_{i,j}) \bbeta_{l} + \br_i(t_{i,j}) \bb_i , \sigma_l^2$) 
		\STATE  $d_{i,j} \gets \mathrm{Normal}(  (1,y_{i}(t_{i,j}), \bx_{i} ) \bbeta_d, \sigma_d^2)$ 
		\STATE $j \gets j+1$ \\ 
		\ELSE 
		\STATE $J_{i} \gets j-1$,  continue $\gets \textbf{false} $
		\STATE $\be_{i,T_i} \gets \{(t_{i,0},d_{i,0}), \dots, (t_{i,J_i},d_{i,J_i})\}$ and $\by_i \gets (y_{i}(t_{i,1}),\dots,y_{i}(t_{i,J_i}))$
		\STATE Compute $  R_i $
		\ENDIF
		\ENDWHILE
	\end{algorithmic}
\end{algorithm}

Next we  compute the gradient of the log-likelihood of the MTPP, $\nabla_{\btheta} \log p(\be_{i, T_i} \mid \by_i,\bx_i,\bphi,\btheta)$, using the parametrization defined in \eqref{eq:dos_vis_like}. The details are described in Supplementary Section D.  Lastly, we integrate out $\bphi$ in computing $\nabla_{\btheta} G_i(\btheta)$ using the Monte Carlo method since it is analytically intractable. 
Suppose that we have $K$ MCMC draws from the posterior distribution of $\bphi$ and we denote the $k$-th draw as $\bphi^{(k)}$, 
then $\nabla_{\btheta} G_i(\btheta)$ can be approximated as follows:
\begin{eqnarray}
\nabla_{\btheta} G_i(\btheta)\approx \frac{\sum_{k=1}^K  E_{ (\by_i, T_i, \be_{i, T_i})\sim p(\by_i, T_i, \be_{i, T_i}\mid \btheta,\bphi^{(k)} , \bx_i) } [R_i \nabla_{\btheta} \log p(\be_{i, T_i} \mid \by_i,\bx_i,\bphi^{(k)}, \btheta) ]}{K}. \nonumber\\
\label{eq:mcmethod}
\end{eqnarray}
To compute each term of the summation in the numerator of \eqref{eq:mcmethod}, we first sample $T_i$, $\by_i$, and $\be_{i, T_i}$  from $p(\by_i, T_i, \be_{i, T_i}\mid \btheta, \bphi^{(k)}, \bx_i)$ using Algorithm \ref{sample_alg} to compute $R_i$ for each $\bphi^{(k)}$, then multiply the gradient of the 
log-probabilities of visitation times and dosages under the MTPP policy.  
We use an adaptive step size algorithm and choose the step size to be: 
$s_{i,m}=\frac{0.01}{ \sqrt{\sum_{l=m-50}^{m-1}\nabla_{\btheta} G_i(\btheta_{i,l})^2} }$.
The entire SGD algorithm for finding the optimal parameter $\widetilde{\btheta}_i$   is described in Algorithm \ref{SGD_alg}, where $\overline{G_i(\btheta_{i,m})}$ denotes the expected reward in iteration $m$.  We select the optimal policy $\widetilde{\btheta}_i$ to be the one yielding the highest expected reward across all  iterations. 
Note that, in the step 7 of Algorithm \ref{SGD_alg}, we subtract the average reward from each individual reward: this ``baseline subtraction'' trick significantly reduce the variance while still yielding an unbiased estimate of the gradient 
\citep{williams1992simple,greensmith2004variance}. 

%

\begin{algorithm}[ht!]
	\caption{Stochastic Gradient Descent for optimizing $\btheta$ for any patient $i$} 
	\label{SGD_alg}
	  \textbf{Input} $\btheta_0$, $\bphi^{(k)}$ ($k=1,\dots K$), $\bx_{i}$, $y_{i,0}$. \\
 \textbf{Output} $\widetilde{\btheta}_i$
	\begin{algorithmic}[1]
		\STATE Initialize $\btheta_{i,1} \gets \btheta_0$ 
		\FOR{m:=1 \TO M-1}
		\FOR{k:=1 \TO K}
		\STATE do Algorithm \ref{sample_alg}($\btheta_{i,m}$, $\bphi^{(k)}$, $\bx_{i}$, $y_{i,0}$) to sample $\be_{i,T_{i}^{(k)}}^{(k)}$ and $\by_{i}^{(k)}$, and compute $R_{i}^{(k)}$.
		\ENDFOR\\
		\STATE $ \overline{G_i(\btheta_{i,m})} \gets  \frac{\sum_{k=1}^K  R_{i}^{(k)} }{K}$ 
		\STATE $\nabla_{\btheta} G_i(\btheta_{i,m})\gets \frac{\sum_{k=1}^K  (R_{i}^{(k)} - \overline{G_i(\btheta_{i,m})} ) \nabla_{\btheta} \text{log}  p(\be_{i,T_{i}^{(k)}}^{(k)} \mid \by_i^{(k)},\bx_i,\bphi^{(k)},\btheta_{i,m})  ) }{K}$\\
		\STATE $\btheta_{i,m+1} \gets \btheta_{i,m} + s_{i,m} \nabla_{\btheta} G_i(\btheta_{i,m}) $
		\ENDFOR\\
		\STATE $ m^* \gets \argmax_m \overline{G_i(\btheta_{i,m})}$ \\ 
		\STATE $ \widetilde{\btheta}_i \gets \btheta_{i,m^*}$
	\end{algorithmic}
\end{algorithm}\texttt{}

\section{Simulation Study}
\label{sec:sim_study}


To demonstrate the advantage of the proposed Bayesian joint model, we compared it to an alternative model that breaks the connection  between  longitudinal and survival processes. Furthermore, to illustrate the benefit of optimizing the  personalized clinical decision, we compared the expected reward under the estimated optimal clinical decision to alternative strategies of scheduling follow-up visitations on a regular basis, e.g., every three months \citep{israni2014variation}.

\subsection{Simulation setup}

We simulated a dataset mimicking the DIVAT dataset composed of longitudinal creatinine measurements, follow-up schedules, tacrolimus dosages, and survival events for $I$ = 500 patients. 
We considered three baseline covariates in $\bx_{i}$: donor age (AgeD), delayed graft function (DGF), and body mass index (BMI). DGF is a binary variable with 1 indicating that the patient used dialysis within the first week of the transplant, 0 otherwise. For each patient, the donor age and BMI were generated from $\mathrm{Normal}(52.5,15.8^2)$ and $\mathrm{Normal}(24.3,4.5^2)$, respectively, and then standardized. Patients' delayed graft functions were generated from $\mathrm{Bernoulli}(0.4)$ independently. In the MTPP model for follow-up schedules, the simulated true parameters were set to be $\nu_1=2.5$, $ \nu_2=1.5 $, $\mu=-4.8$, $\xi=2$, and $\bbeta_\alpha=(9.5,-1.5)^T$ so that a higher creatinine level results in a higher visitation intensity; for assigning dosages, the simulated true $\bbeta_d$ was set to be $(1,0.2,0.15,0.2,0.15)^T$ and $\sigma_d=0.3$.  
In modeling log-transformed longitudinal creatinine levels,  the simulated true parameters were set to be $\bbeta_l= (5.3,0.1,0.3,0.4, 0.25, -1 \times 10^{-4},3\times 10^{-8} )^T$,  $\sigma_l=0.1$, and $\bSigma_b= \begin{bmatrix}
0.04 & 0 & 0 \\
0    & 0.0049 & 0\\
0    &    0   & 10^{-8}\\
\end{bmatrix}
$. Note that the last two terms in the simulated true $\bbeta_l$ were small since the times were recorded in days. 
Patients' initial log-transformed creatinine levels right after transplantation $y_{i,0}$'s were independently generated from $\mathrm{Normal}(5,0.1^2)$. 
In the survival submodel \eqref{eq:surv_haz}, we assumed that the simulated true parameters were  
$h_0=5$, $\omega=1.05$, $\beta_{s1}=1$,    $\beta_{s2}=0.9 $, $\beta_{s3}=-0.75$, $\beta_{s4}=-5$, and $\eta_{tox}=50$. The censoring times $C_i$'s were independently generated from  $ \mathrm{Weibull}(3,8000) $. Based on the proposed Bayesian joint model in Section \ref{sec:joint}, we generated the data  $\by_i,\be_{i, \widetilde{T}_i}, \widetilde{T}_i,\delta_i$ for each patient $i$, $i=1, \dots, I$. 


The simulated dataset had a total of 14,395 follow-up visitations for $500$ patients with a 10.8\% censoring rate. The median survival time was 1,684 days with the shortest being 24 days and the longest being 10,016 days. Supplementary Figure S1 plots the simulated longitudinal creatinine levels and follow-up schedules with dosages for four randomly selected patients.

\subsection{Results: model fitting}
We applied the proposed Bayesian joint model to the simulated dataset. The hyperparameters were set to be $\bbeta_{d0}=\bbeta_{l0}=\bbeta_{\alpha 0}=\bf{0}$, $\bSigma_{\bbeta_{d}}=\bSigma_{\bbeta_{l}}=\bSigma_{\bbeta_{\alpha}}=100^2 I$, $\pi_{d1}=\pi_{d2}=\pi_{l1}=\pi_{l2}=\pi_{s3}=\pi_{s4}=0.01$, $\pi_{s1}=\pi_{s2}=0.01$, $\beta_{s0}=\beta_{v0}=0$, $\sigma_{s0}^2=\sigma_{v0}^2=100^2$, $\pi_{v1}=400$, $\pi_{v2}=200$. 
We ran 20,000 MCMC iterations with an initial burn-in of 5,000 iterations and a thinning factor of 50. The convergence was assessed using R package {\it coda}, including traceplots of the post-burn-in MCMC samples for some randomly selected parameters (Supplementary Figure S2), showing no issues of non-convergence.  We first report on the performance of the proposed joint model in terms of parameter estimation. Supplementary Figure S3 plots the 95\% estimated credible intervals (CIs) for selected parameters, showing that all  95\% CIs are centered around the simulated true values. As another metric of performance, we computed the mean squared error (MSE) taken as the averaged squared errors between
the post-burn-in MCMC posterior samples and the simulated true values. Supplementary Table S1 summarizes the MSE and the standard deviation of squared errors, indicating that the proposed joint model can accurately estimate parameters. 

As the proposed model represents the first effort in the literature to jointly model clinical decisions, longitudinal markers, and the survival event, there is no existing method we can compare with. To demonstrate the advantage of jointly modeling longitudinal creatinine levels and the survival event, we compared the proposed model with an alternative ``separate longitudinal and survival (SLS)'' model that breaks the connection between the longitudinal and survival submodels by 
replacing the process $y_{i}^*(t)$ with the observational data $y_{i}(t)$ in the hazard model \eqref{eq:surv_haz}.  
We first compared the two models by checking their model adequacy using the Watanabe-Akaike information criterion  (WAIC) \citep{watanabe2010asymptotic}: the joint model has a WAIC value of 226,982 while the SLS model has a WAIC of 226,992, indicating that the proposed joint model fits data slightly better. Furthermore, we compared the two models in terms of parameter estimation. Table \ref{Tab:param_sim_est} reports the simulated true values of parameters in the survival submodel, and posterior means of these parameters under the joint model and the SLS model with 95\% CIs, showing that the joint model estimates parameters more accurately. 

\begin{table}[H] 
	\centering
	\begin{tabular}{|c|c|c|c|}
		\hline
		& Truth & Joint posterior mean (95\% CI) & SLS posterior mean (95\% CI) \\ 
		\hline
		$\beta_{s1}$ & 1 & 1.1(0.92,1.26) & 1.19(0.94,1.6) \\ 
		\hline
		$\beta_{s2}$ & 0.9 & 1.25(0.74,1.95) & 1.41(0.63,2.3) \\ 
		\hline
		$\beta_{s3}$ & -0.75 & -0.92(-1.62,-0.33) & -1.03(-1.8,-0.18) \\ 
		\hline
		$\beta_{s4}$  & -5 & -5.01(-5.51,-4.47) & -5.16(-6.14,-4.56) \\ 
		\hline
		$h_0$ & 5 & 4.36(3.44,5.35) & 3.89(1.6,5.22) \\ 
		\hline
		$\omega$ & 1.05 & 1.06(0.99,1.12) & 1.06(0.97,1.13) \\ 
		\hline
	\end{tabular}
	\caption{Parameter estimation under the joint and SLS models. }
	\label{Tab:param_sim_est}
\end{table}

%
\subsection{Results: personalized optimal clinical decision estimation}
We applied the proposed policy gradient method in Section \ref{sec:SGD} to the simulated dataset to estimate the personalized optimal clinical decision that maximizes one patient's graft median survival time, i.e., $R_i=\log(\widehat{T}_i)$, where $\widehat{T}_i$ is the median survival time of any patient $n$.  The starting parameter values $\btheta_0$ in Algorithm \ref{SGD_alg} were set to be the estimated posterior means of these parameters from posterior inference, which  can be considered as the estimates of how physicians treated patients in the simulated data. Therefore, the goal of the optimization procedure is to improve physicians' current treatment strategy  in terms of prolonging patients'  survival. 

\begin{figure}[ht!]
	\centering 
	\begin{tabular}{cc}
		\includegraphics[page=2,scale=0.35]{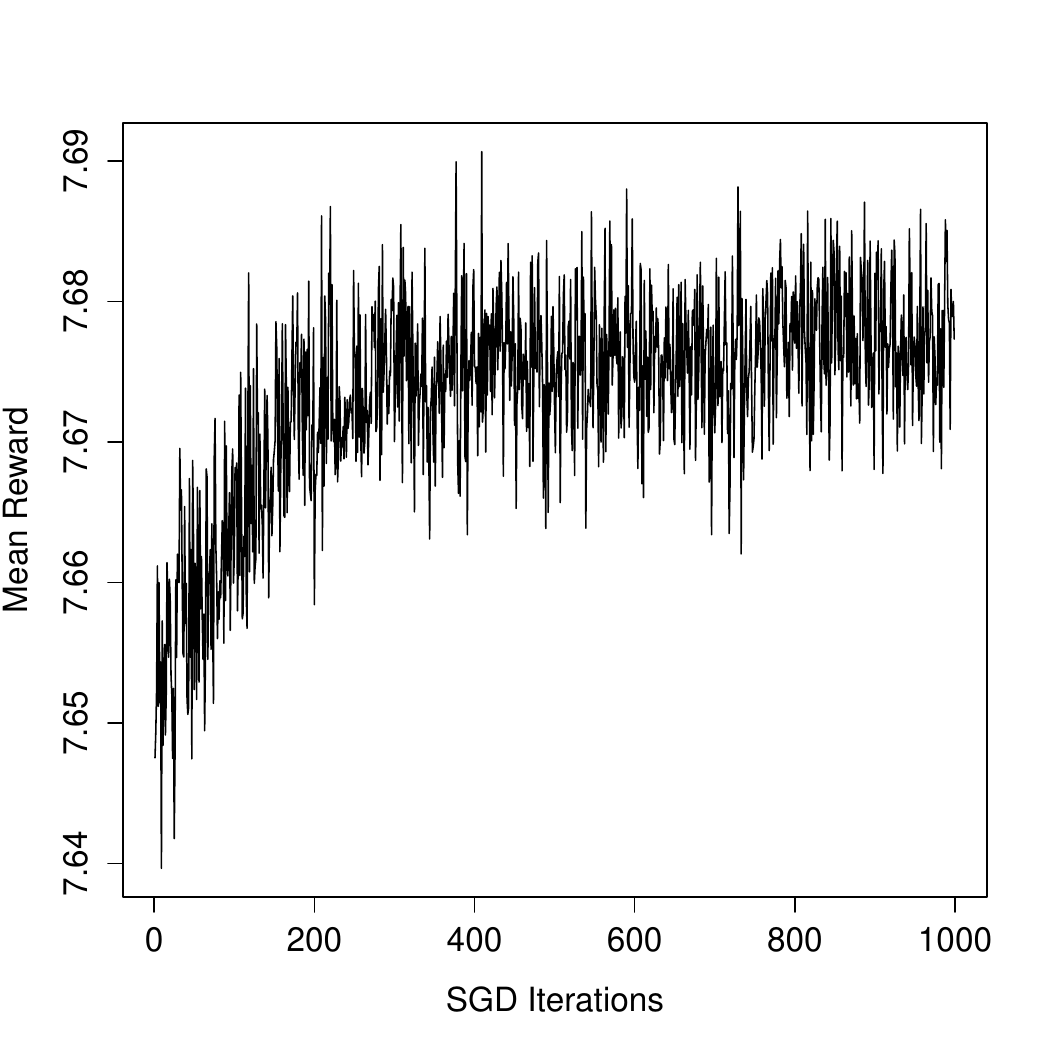} & 
		\includegraphics[page=2,scale=0.35]{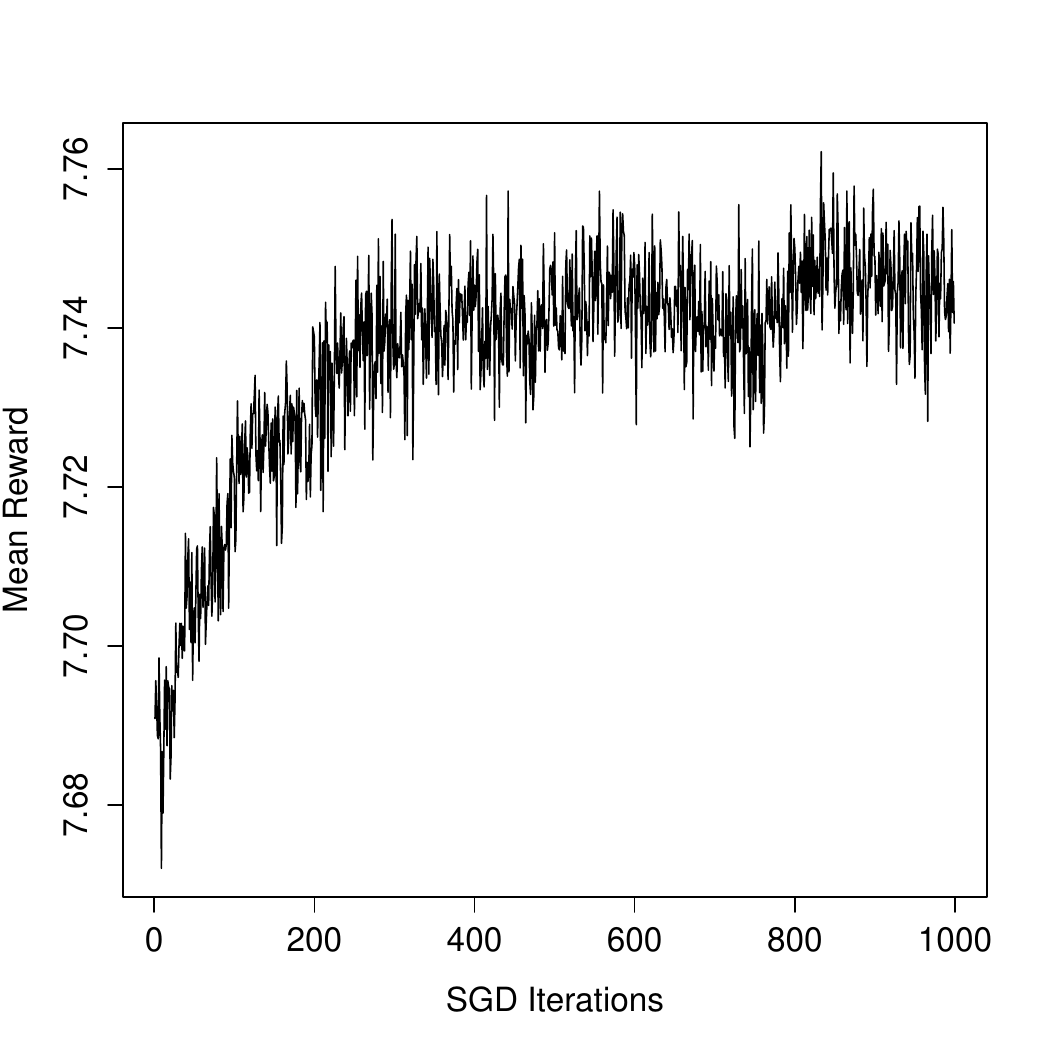}\\
		(a) Patient S1 & (b) Patient S2 \\
		\includegraphics[page=2,scale=0.35]{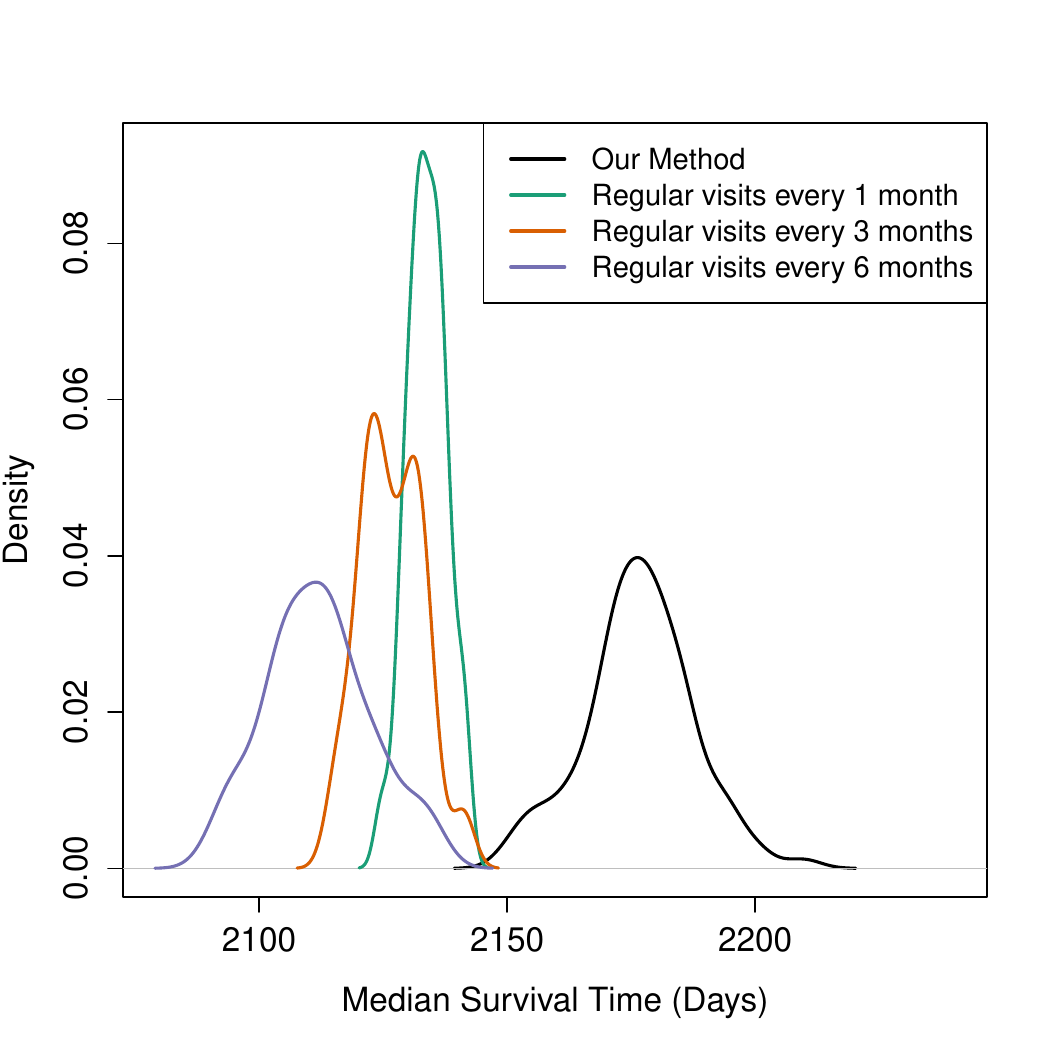} & 
		\includegraphics[page=4,scale=0.35]{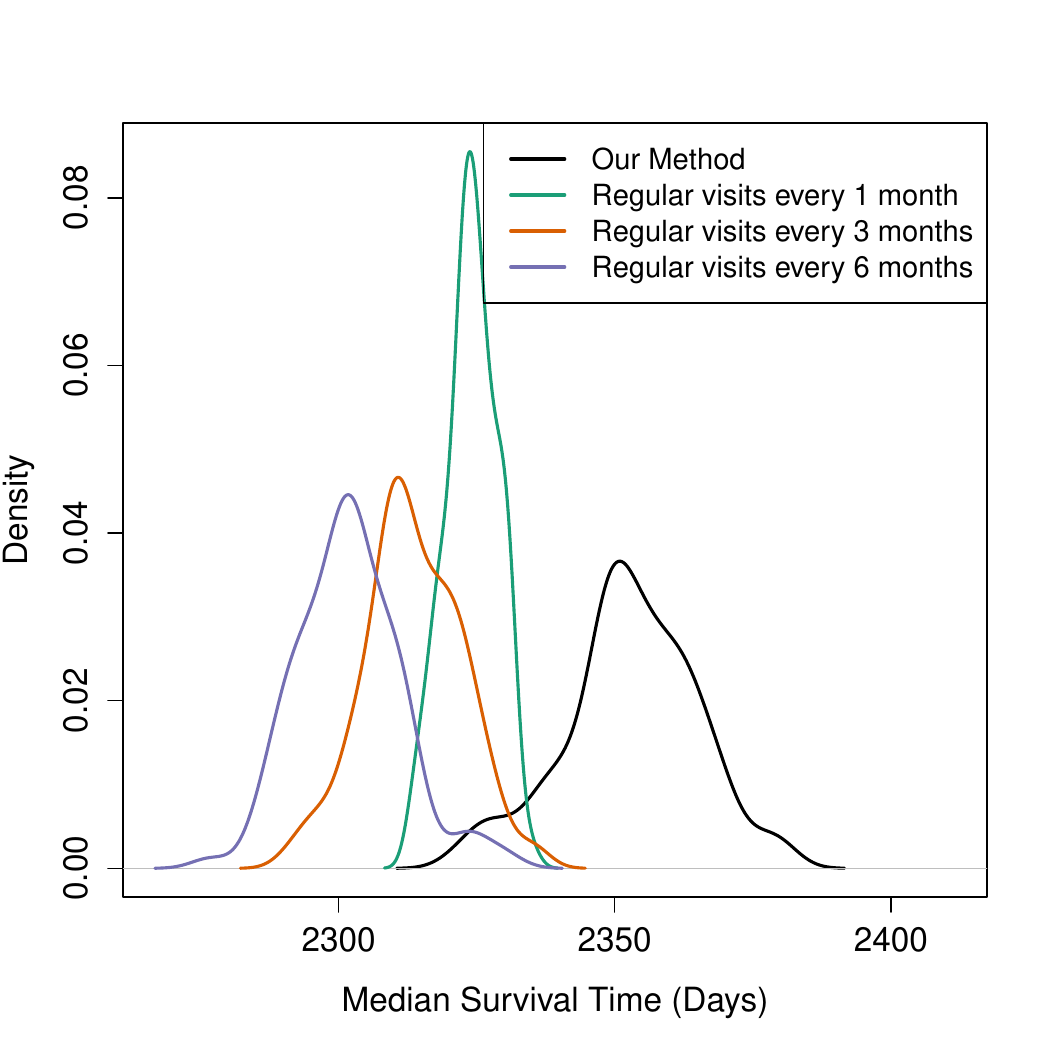}\\
		(c) Patient S1 & (d) Patient S2 \\
	\end{tabular}
	\caption{Panels (a, b) plot the expected mean reward versus SGD iterations for two randomly selected patients S1 and S2. Panels (c, d) plot the density of the predictive median survival times under our method and the three alternative strategies for patients S1 and S2.}
	\label{fig:grad_result_sim}
\end{figure}

We implemented Algorithm \ref{SGD_alg} with $M=1000$ steps to estimate the personalized optimal parameter  $\widetilde{\btheta}_i$ for two randomly selected patients, denoted as S1 and S2. Patient S1 had a DGF of 0, donor age of 54.2 years, and BMI of 24, while patient S2 had a DGF of 1, donor age of 37.4 years, and BMI of 24.8. Figure  \ref{fig:grad_result_sim}(a, b) plots the expected mean reward versus SGD iterations. Note that the SGD procedure in the context of reinforcement learning usually has high variability, which is alleviated by our variance reduction method in Algorithm \ref{SGD_alg} as shown in Figure  \ref{fig:grad_result_sim}. 
For patient S1, the expected mean reward  increases from an initial value of 7.65 to its maximum in the SGD, 7.69, 
which corresponds to a predictive median survival time of 2,209 days, a 4.6\% increase from its initial value 2,111. For patient S2, the expected mean reward goes from an initial value of 7.69 to a maximum at 7.76. 
This corresponds to the predictive median survival time increasing from 2,203 days to 2,383 days, an 8.2\% improvement.

To further interpret the estimated optimal ``policy" parameters  for  patients S1 and S2, we compared the initial parameter values of the SGD--posterior means, with the optimized values by the SGD in Table \ref{Tab:param_sim}. Recall that the dosage model is 
$d_{i,j}=  (1,y_{i,j}, \bx_{i} ) \bbeta_{d}  + \zeta_{i,j}$. Denote $\bbeta_d=(\beta_{d1}, \beta_{d2}, \dots, \beta_{dL})^T$, where $L$ is the dimension of $\bbeta_d$.
Since $\bx_i$ denotes the baseline covariate and does not change over time, we define the personalized dosage intercept to be $\tilde{\beta}_{d}=(1, \bx_i)(\beta_{d1}, \beta_{d3}, \dots, \beta_{dL})^T$ so that optimizing $\bbeta_d$ is equivalent to optimizing ($\beta_{d2}, \tilde{\beta}_d$). 
As shown in Table \ref{Tab:param_sim}, the optimized dosage parameters $\tilde{\beta}_{d}$ and $\beta_{d2}$  for patient S1 were lower than the estimated posterior means, indicating that patient S1 would benefit from a lower dosage for the same creatinine level compared to the observed dosages. In contrast, the optimal $\tilde{\beta}_{d}$ and $\beta_{d2}$ were higher than the posterior means for patient S2, indicating the preference for higher dosages. The optimal dosage errors, $\sigma_d^2$, for both patients were significantly lower than the initial value, indicating that a lower variance in the dosing procedure would benefit patient survival. The optimal baseline visitation intensity $\mu$ and the peak time parameter $\nu_1$ were both roughly the same as their posterior means, indicating that the simulated follow-up schedules were close to optimal. However, the visitation intensity shape parameter $\nu_{2}$  increased from 1.464 to 1.778 and 2.008 for patients S1 and S2 respectively and thus implies a higher intensity around the peak time $\nu_{1}$: intuitively,   the optimized policy learns to be more certain about the ``optimal peak time.''  

\begin{table}[ht!] 
	\caption{Simulation: Stochastic Gradient Descent Optimal Parameter Results}
	\centering
	\begin{tabular}{ |l|c|c|c| } 
		\hline
		& $\widetilde{\btheta}_{0}$ &  $\widetilde{\btheta}_{S1}$  & 
		$\widetilde{\btheta}_{S2}$   \\ 
		\hline
		$\tilde{\beta}_{d}$: personalized dosage intercept & S1: 0.864, S2:0.987  &  0.746 & 1.316 \\ 
		\hline
		$\beta_{d2} $: dosage effect of creatinine & 0.200 & 0.153 & 0.307 \\ 
		\hline 
		$\sigma_d^2 $: dosage error & 0.0940 & 0.0217 & 0.00252 \\ 
		\hline 
		$\mu$: baseline visitation intensity  & -4.781 & -4.821 & -4.785 \\ 
		\hline
		$\nu_1$: visitation intensity peak & 2.512 & 2.416 & 2.519 \\ 
		\hline
		$\nu_2$: visitation intensity shape & 1.464 & 1.778 & 2.008 \\ 
		\hline
	\end{tabular}
	\label{Tab:param_sim}
\end{table}

In addition, to illustrate the advantage of optimizing both follow-up  schedules and dosages, we compared our results to alternative strategies based on regular visits. As studied in \cite{israni2014variation}, during the first year post-transplant, patients were most frequently seen every 1 month or 3 months, depending on their physicians. After the first year, stable patients were most frequently referred back between 4-6 months but the follow-up  frequency was reported to vary from 0-12 months. We considered three alternative follow-up strategies: recommend patients to follow up every 1 month, 3 months, and 6 months. The dosages at follow-up visitations were still optimized in the same way as the proposed joint model with the policy gradient method. Figure \ref{fig:grad_result_sim}(c, d) show the density plots of 100 realizations of the predictive median survival times under our method and the three alternative strategies for patients S1 and S2. Comparing the predictive median survival times under the three regular visitation strategies, we can see that more frequent visitations yield longer median survival times. The optimized visitation schedule under the proposed method outperforms the three alternative strategies although it yields a similar overall visitation frequency with the strategy of ``regular visits every 3 months" (not shown),  
highlighting the importance of optimizing   visitation schedules based on  longitudinal clinical measurements  to prolong patients' survival. 

\section{Application: DIVAT Data Analysis}
\label{sec:realdata_analysis}
We extracted data from Nantes University Hospital Centers in the DIVAT cohort (www.divat.fr), yielding 
a total of $I=947$ patients who received a first or second renal graft transplanted from a living or heart-beating deceased donor between 2000 and 2014. All patients in the dataset received an initial maintenance therapy with tacrolimus and did not experience graft failure or death during hospitalization. Immediately after transplantation, several baseline covariates as risk factors for graft failure were collected: donor age (AgeD), recipient age (AgeR), delayed graft function (DGF) defined as the indicator of the use of dialysis within the first week of transplant (1=used dialysis, 0=didn't use dialysis), diabetes history (Diab) with 1 indicating the patient has a history of diabetes and 0 otherwise,  type of donor (Type), and body mass index (BMI).   There were two types of donors: donation after brain death but with heart beating (Type=1) and donation by a living donor (Type=0).  Table \ref{table:data} summarizes patients' characteristics at baseline immediately after transplantation. For each patient, longitudinal data were collected from the date of transplantation until the graft failure or being censored.  At each follow-up visitation, the creatinine level and tacrolimus dosage were recorded. The next follow-up visitation time was determined by the physician. 

\begin{table}[h] 
	\begin{tabular}{p{10cm}r}
		\hline
		\multicolumn{2}{l}{Donor age (years)} \\
		\makebox[5cm][c]{Mean $\pm$ SD }  & 52.5 $\pm$ 15.8 \\
		\makebox[5cm][c]{Median }  & 54\\
		\\
		\multicolumn{2}{l}{Receipient age (years)} \\
		\makebox[5cm][c]{Mean $\pm$ SD }  & 51.1 $\pm$ 14.3 \\
		\makebox[5cm][c]{Median }  & 52 \\
		\\
		\multicolumn{2}{l}{Body mass index (BMI)} \\
		\makebox[5cm][c]{Mean $\pm$ SD }  & 24.3 $\pm$ 4.5\\
		\makebox[5cm][c]{Median }  & 23.7\\
		\\
		\multicolumn{2}{l}{Delayed graft function, n(\%)} \\
		\makebox[5cm][c]{Yes} & 329 (34.7\%)\\
		\makebox[5cm][c]{No } & 618 (65.3\%)\\
		\\
		\multicolumn{2}{l}{Diabetes history, n(\%)} \\
		\makebox[5cm][c]{Yes} & 140 (14.5\%)\\
		\makebox[5cm][c]{No } & 807 (85.5\%)\\
		\\
		\multicolumn{2}{l}{Type of donor, n(\%)} \\
		\makebox[5cm][c]{Yes} & 800 (84.5\%)\\
		\makebox[5cm][c]{No } & 147 (15.5\%)\\
		\hline
	\end{tabular}
	\caption{ Patient characteristics at baseline immediately after transplantation. }
	\label{table:data}
\end{table}




\subsection{Experimental results: model fitting}

We first applied the proposed Bayesian joint model to the DIVAT data with $\bx_i= (\mathrm{AgeD}_i,\mathrm{AgeR}_i,\mathrm{DGF}_i,\mathrm{BMI}_i, \mathrm{Diab}_i,\mathrm{Type}_i)$. The hyperparameters were set to the same as in the simulation study. 
We ran a total of 20,000 MCMC iterations with an initial burn-in of 5,000 iterations, and a thinning factor of 50. The convergence was assessed using R package {\it coda} and the trace plots for randomly selected parameters were shown in Supplementary Figure S4, showing no issues of non-convergence. 

\begin{figure}[ht!]
	\centering 
	\includegraphics[page=2,width=1\textwidth]{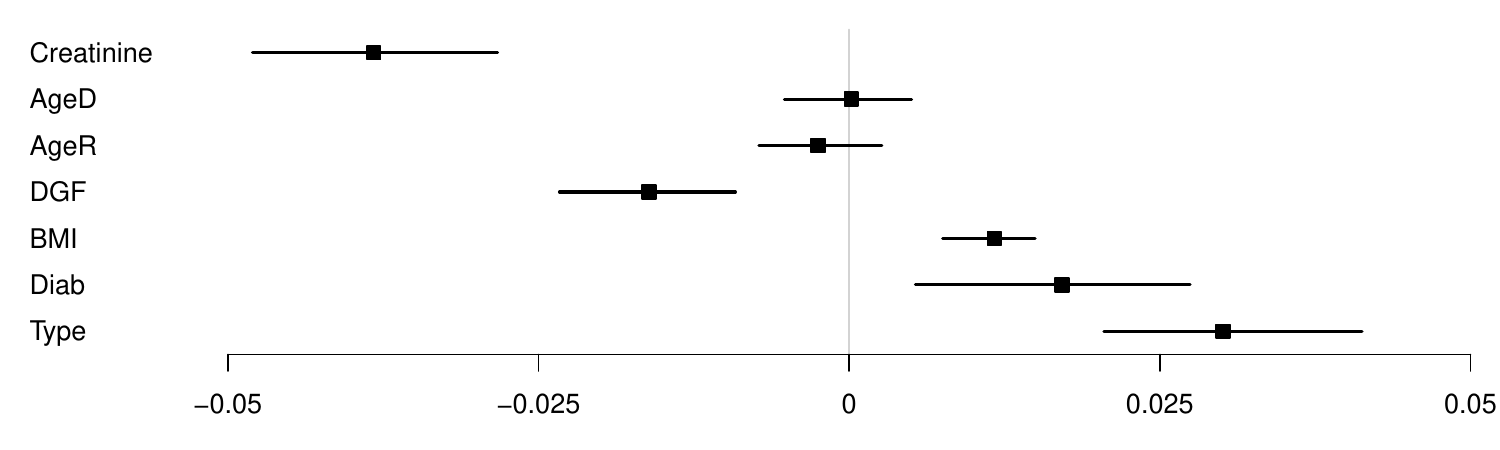} \\
	(a) Dosage parameters\\
	\includegraphics[page=3,width=1\textwidth]{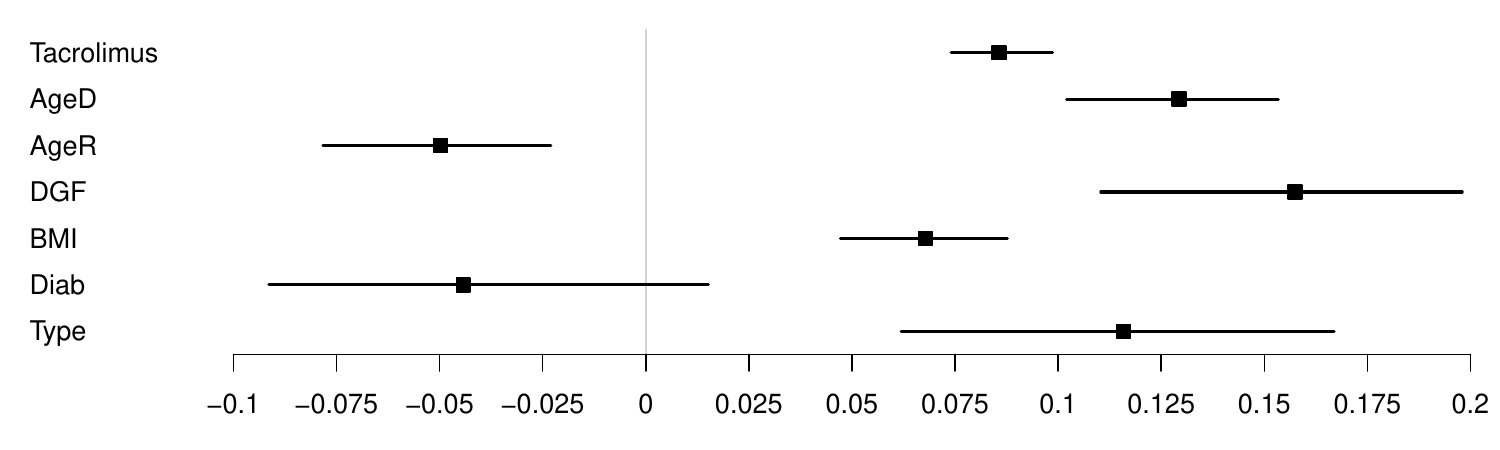} \\
	(b)  Longitudinal parameters\\
	\includegraphics[page=4,width=1\textwidth]{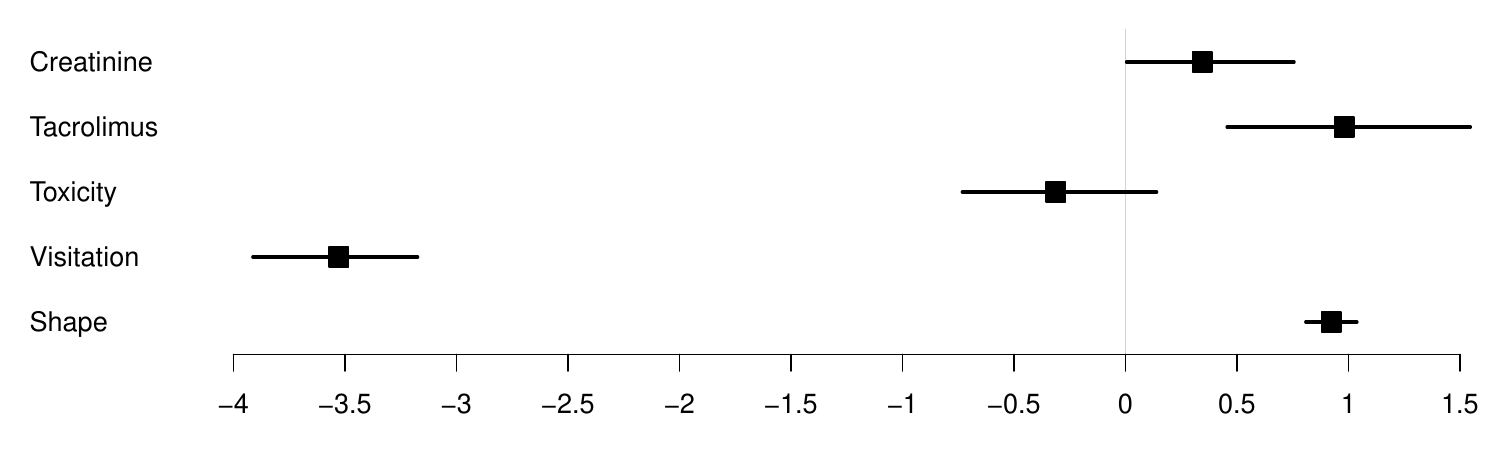}\\
	(c) Survival parameters \\ 
	\caption{Estimated posterior means and 95\% CIs for parameters in the  dosage, longitudinal, and survival submodels. The dosages and longitudinal measurements  are in log-scale. The squares represent  posterior means.}
	\label{fig:cred_real}  
\end{figure}

We plot the estimated posterior means with 95\% CIs for some selected parameters in the dosage, longitudinal, and survival submodels in  Figure \ref{fig:cred_real}. Figure \ref{fig:cred_real}(a) plots posterior means of the linear coefficient $\bbeta_d$ with respect to the creatinine level and baseline covariates in the dosage  model. DGF was negatively associated with the dosage, indicating that patients who used dialysis within the first week of transplant were likely to be assigned lower dosage levels. In contrast, BMI was positively associated with the dosage since bodyweight-based dosing of tacrolimus is the standard care for patients after transplantation \citep{andrews2017overweight}. Diabetes history was positively associated with the dosage. While the effect of diabetes on tacrolimus was not well characterized in the literature, \cite{mendonza2007tacrolimus} showed that the time to maximum concentration of tacrolimus in the pharmacokinetics study was significantly longer in diabetics versus nondiabetics. Furthermore, donor type also increased the dosage level, indicating that patients who received kidney from a non-living donor were  more likely to be assigned higher dosages compared to that from a living donor. 


Figure \ref{fig:cred_real}(b)  plots the estimated posterior means with 95\%  CIs  for  the fixed-effects regression coefficients with respect to the most recent tacrolimus dosage and   baseline covariates in the longitudinal model \eqref{eqn:true_creatinine}. The dosage, donor age, DGF, BMI, and donor type were positively associated with the creatinine level, which agreed with findings in the  literature  \citep{katari1997clinical,gerchman2009body, Foucher2016}. In contrast, the recipient age was negatively associated with the creatinine level, suggesting that younger patients tend to have lower creatinine levels \citep{Maraghi2016}. Diabetes history also decreased the creatinine level.  \cite{hjelmesaeth2010low} showed that a low creatinine was associated with type 2 diabetes in a cross-sectional study.
The estimated posterior means and 95\% CIs for selected survival submodel parameters are plotted in Figure \ref{fig:cred_real}(c). 
The posterior mean of the parameter corresponding to the tacrolimus dosage was positive while that corresponding to the toxicity was negative, suggesting that a higher tacrolimus drug reduces the  hazard but the accumulated toxicity increases the hazard. 
These results were consistent with  findings in \cite{randhawa1997} and \cite{Bottiger1999}, who reported  nephrotoxicity caused by long-term high dosages of tacrolimus. 

\subsection{Experimental results: personalized optimal clinical decision estimation}\label{sec:divat_personal}

\begin{figure}[ht!]
	\centering 
	\begin{tabular}{cc}
		\includegraphics[page=1,scale=0.35]{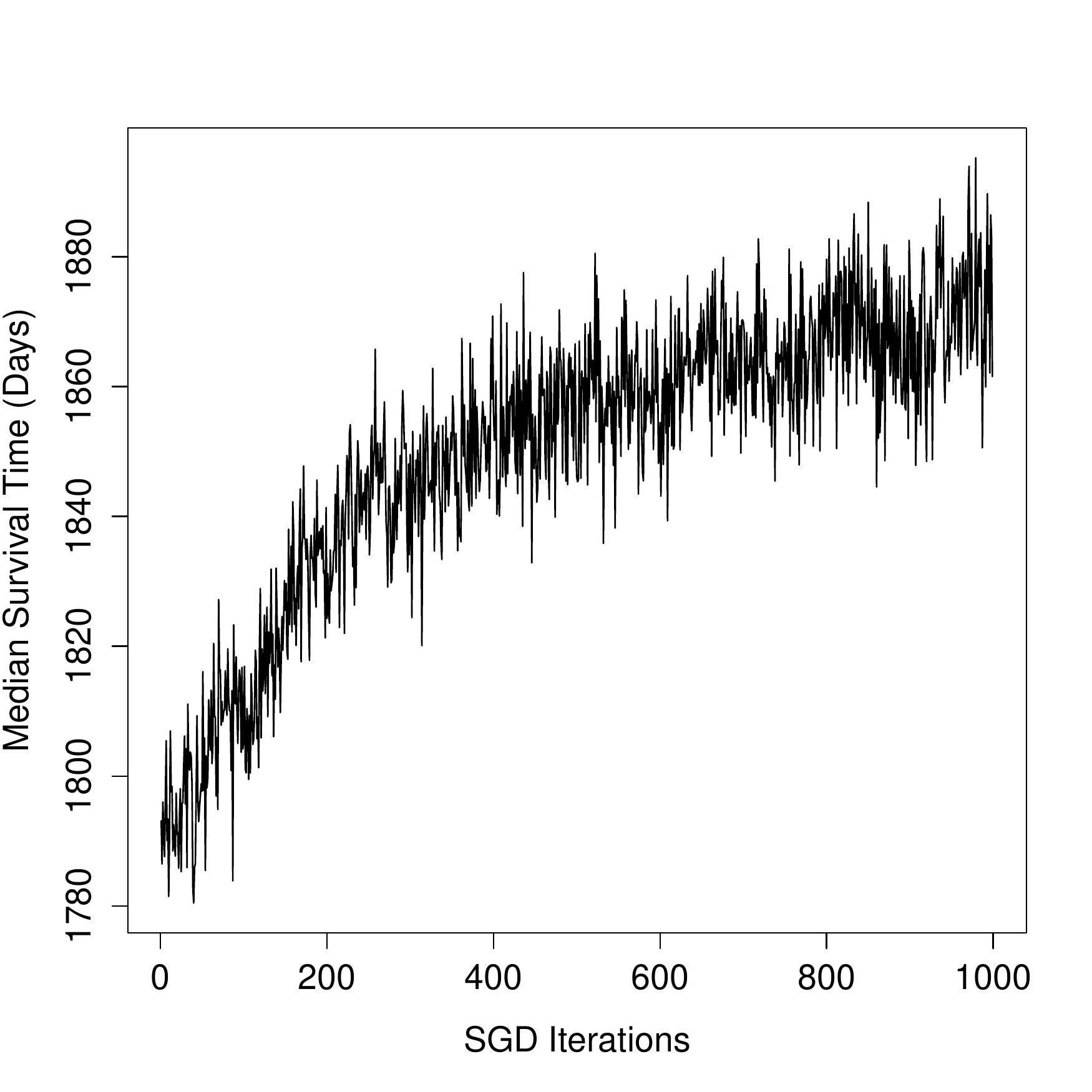} & 
		\includegraphics[page=1,scale=0.35]{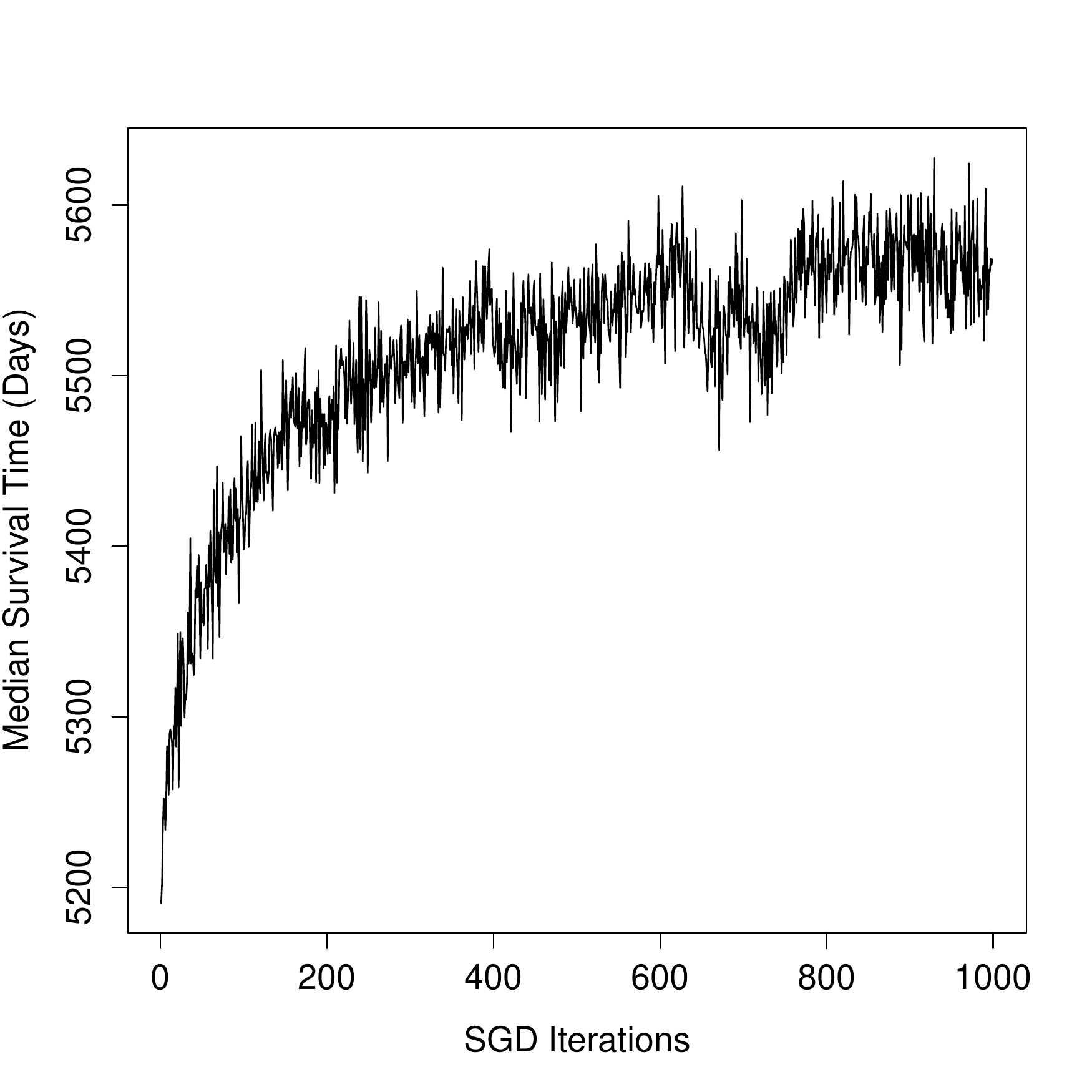}\\
		(a) Patient R1 & (b) Patient R2 \\
	\end{tabular}
	\caption{The expected mean reward versus SGD iterations for two randomly selected patients R1 and R2. }
	\label{fig:grad_result_real1}
\end{figure}

Next, we applied the proposed policy gradient method to estimate the personalized optimal clinical decision in terms of maximizing a patient's median survival time. We initialized the parameters in Algorithm \ref{SGD_alg} by setting $\btheta_0$ to be their posterior means. Algorithm \ref{SGD_alg} was implemented with $M=1000$ steps to estimate   $\widetilde{\btheta}_i$ for two randomly selected patients, denoted as R1 and R2. Patient R1 at transplantation was 60 years old with a BMI of 17, no history of diabetes, no DGF, and received donation from a 61-year-old non-living donor. Patient R2 at transplantation was 28 years old with a BMI of 25.5, no history of diabetes, no DGF, and received a kidney from a living 29-year-old donor. Patient R1 had an observed survival time of 1,527 days, while patient R2 had a censored survival time of 4,487 days. Figure \ref{fig:grad_result_real1} plots the predictive median survival times across SGD iterations for the two patients. Patient R1's predictive median survival time increased from 1,793 to 1,895 days at the maximum, a 5.7\% improvement; while
patient R2's predictive median survival time increased from 5,191 to 5,628, an 8.4\% gain. 


To further interpret the estimated optimal parameters in  clinical decisions, we compared their initial values with the optimized values in Table \ref{Tab:param_real}. In practice, these optimized  parameters  can  be  interpreted  to  guide  clinical  decisions. 
Patient R1's optimal dosage parameters, $\tilde{\beta}_{d}$ and $\beta_{d2}$, were higher than their posterior means, suggesting that 
assigning a higher dosage level compared to what the physician actually did for the same creatinine level would improve his/her survival outcome. Specifically, the recommended dosage for patient R1 based on the optimized $\tilde{\beta}_{d}$ and $\beta_{d2}$ is the baseline of 2.788 plus an additional 0.076 units of tacrolimus for each unit of creatinine recorded at the most recent followup visitation. 
On the other hand, patient R2's optimal dosage parameters were both lower than the initial values, so lower dosage levels are recommended. 
The optimal dosage errors, $\sigma_d^2$, for both patients were significantly lower than the initial value, meaning that the optimized policy is more certain about its dosing decisions so the variance is lower than the observed data. Therefore, a decrease in the optimized dosage error term can be implemented in practice by closely adhering to the mean dosage amount suggested by the optimized dosage parameters.
The optimal baseline visitation intensities $\mu$ for both patients were lower than the initial value, indicating that they should  be instructed to visit less often without the knowledge of their creatinine measurements.   
Their optimized visitation intensity peak times were lower than the posterior mean, indicating that they should be scheduled more frequent follow-ups when their creatinine levels are high. Furthermore, the visitation intensity shapes were significantly higher than the initial value so the optimized policy is more certain about the optimal peak time for visitation schedules. 

\begin{table}[ht!] 
	\caption{DIVAT data:  optimal parameters  estimated by the policy-optimizing method. }
	\centering
	\begin{tabular}{ |l|c|c|c| } 
		\hline
		& $\widetilde{\btheta}_{0}$ &  $\widetilde{\btheta}_{R1}$  & 
		$\widetilde{\btheta}_{R2}$   \\  
		\hline
		$\tilde{\beta}_{d}$: personalized dosage intercept & R1:2.367, R2:2.363 & 2.788 & 2.161 \\ 
		\hline
		$\beta_{d2}$: dosage effect of creatinine & -0.038 & 0.076 & -0.065 \\ 
		\hline 
		$\sigma_d^2$: dosage error & 0.111 & 0.035 & 0.0024 \\ 
		\hline 
		$\mu$: baseline visitation intensity  & -4.197 & -4.617 & -4.322 \\ 
		\hline
		$\nu_1$: visitation intensity peak & 1.479 & 1.123 &  1.311  \\ 
		\hline
		$\nu_2$: visitation intensity shape & 0.258 & 0.864 & 1.261 \\ 
		\hline
	\end{tabular}
	\label{Tab:param_real}
\end{table}

\subsection{Ablation study: optimizing time or dosage or both}

\begin{figure}[ht!]
	\centering 
	\begin{tabular}{cc}
		\includegraphics[page=1,scale=0.35]{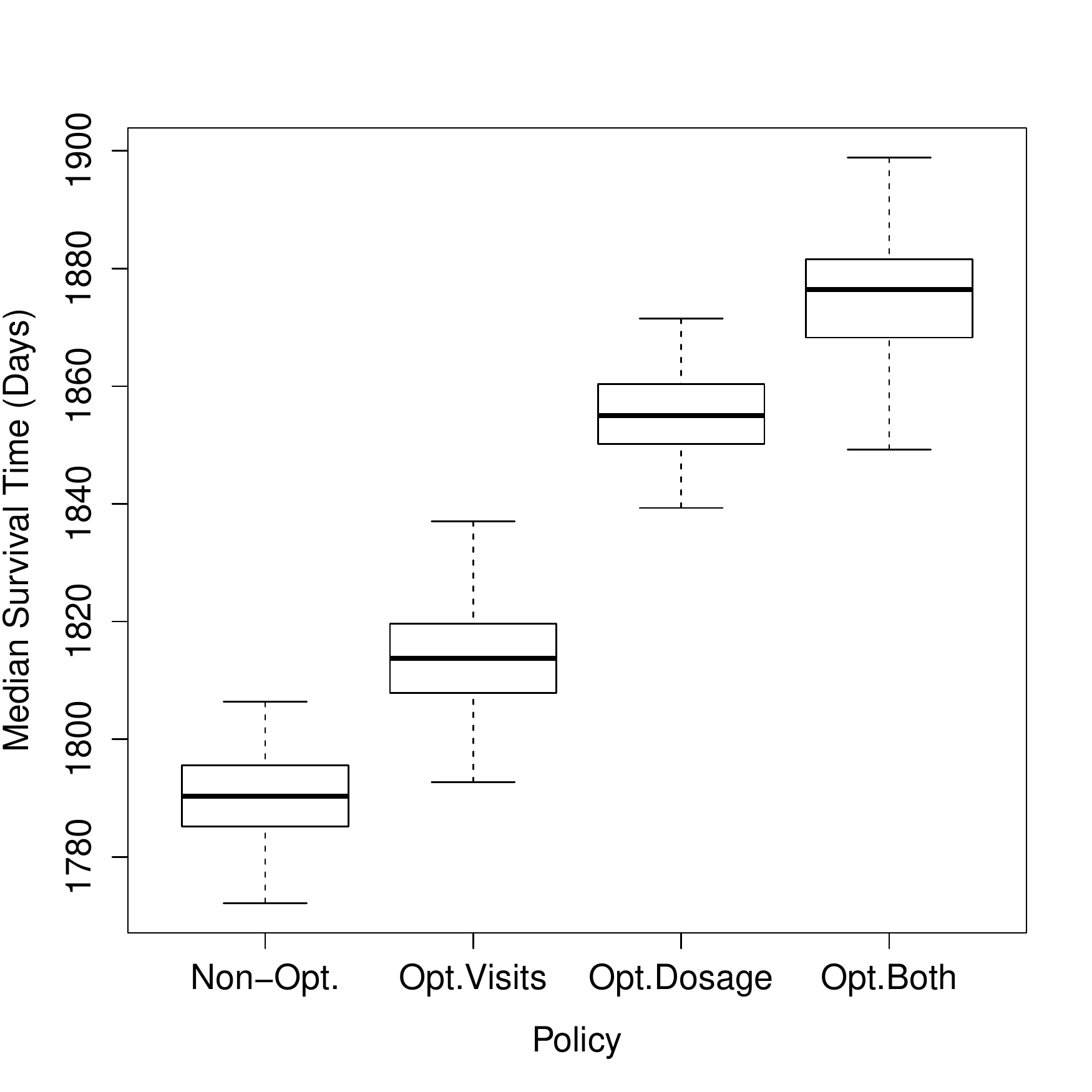} & 
		\includegraphics[page=2,scale=0.35]{Pred_opt_vis_dosage_real.pdf}\\
		(a) Patient R1 & (b) Patient R2 \\
	\end{tabular}
	\caption{The boxplots of the predictive median survival times under different policies of visitation schedules and dosages  for patients R1 and R2.}  
	\label{fig:grad_result_real2}
\end{figure}

Moreover, to demonstrate the benefit of optimizing the follow-up visitation schedules and dosages together, we compared the predictive median survival times under the non-optimized initial policy (Non-Opt.) with three versions of optimized policies: 1) only visitation schedules are optimized (Opt.\@ Visits); 2) only dosages are optimized (Opt.\@ Dosage); and 3) both visitation schedules and dosages are optimized (Opt.\@ Both). Specifically, Non-Opt. used the parameters estimated from the proposed Bayesian joint model, mimicking what physicians did   as collected    in the DIVAT dataset; Opt.\@ Visits used the optimized parameters from the SGD in the visitation model \eqref{eqn:intensity} and the non-optimized parameters in the dosage model \eqref{eqn:dosage}; Opt.\@ Dosage used the optimized parameters from the SGD in the dosage model  and the non-optimized parameters in the visitation model; Opt.\@ Both   is the fully optimized model obtained in \cref{sec:divat_personal} which  used the optimized parameters in both the visitation and dosage models. 
Figure \ref{fig:grad_result_real2} plots   boxplots for 100 realizations of the predictive median survival times under each of the four policies. The visitation schedule optimization accounts for more improvement in prolonging the  survival  for patient R1 compared to patient R2 because, as shown in \cref{Tab:param_real},   there was a larger difference between the optimal parameter values ($\mu$ and $\nu_1$) in the visitation model and their initial values for patient R1. 
The optimized visitation schedule for both patients, as we have discussed in \cref{sec:divat_personal},  suggested slightly fewer visits overall, but more frequent visits when their creatinine levels are high. 
Comparing Opt.\@ Visits vs.\ Non-Opt.\ and Opt.\@ Both vs.\ Opt.\@ Dosage, we can see that optimizing treatment schedules is clearly beneficial to these patients, thus empirically strengthening the motivation of our work. 
In summary, this analysis reveals that optimizing both visitation schedules and dosages is necessary to maximize patients' survival.  

\section{Conclusion}
\label{sec:conclusion}

In this work, we developed a  two-step  Bayesian  approach  to  optimize  clinical  decisions  with timing. Firstly, we proposed a Bayesian joint model for clinical observations (e.g, longitudinal  measurements and survival time) and clinical decisions (e.g., follow-up visitation schedules and dosage assignments). 
The model components are connected by sharing certain structures and parameters in order to capture the mutual influence between the clinical observations and decisions. 
Moreover, we proposed a policy gradient method that optimized the personalized clinical decision for better survival, 
while parameter uncertainties in the clinical observation model are considered.
Through simulation studies, we   demonstrated that the optimized clinical decision obtained from the proposed approach yields longer predictive median survival times compared to scheduling follow-up visitations on a regular basis that is commonly used in caring for patients with chronic conditions nowadays. The analysis of the DIVAT data yields meaningful and interpretable results, showing that the proposed method has the potential to assist physicians’ decisions on personalized treatment. In addition, we have built an R package {\it doct} so that users can apply the proposed method to datasets in a similar setup that involves longitudinal decision making and an objective reward to optimize. 

There are several potential extensions. Firstly, we consider one longitudinal measurement in the longitudinal process of the joint model. There could be other time-varying measurements affecting the clinical decision and survival.  In our kidney transplantation application, besides creatinine levels, there are other longitudinal measurements recorded such as  proteinuria, which represents having  protein in the urine and can be an early sign of kidney disease. The proposed method can be extended to incorporate other longitudinal measurements by replacing the model in \eqref{eqn:true_creatinine} with a multivariate mixed effects model \citep{chi2006joint}. 
Secondly, the proposed Bayesian joint framework that models both clinical decisions and observations relies on certain parametric assumptions that are suitable for our kidney transplantation application. However, these assumptions (e.g., the proportional hazard assumption in the survival submodel) may not hold for other medical applications. It will be straightforward to extend the proposed Bayesian model to more flexible models such as replacing the current prior distributions with a Bayesian nonparametric prior (e.g., the Dirichlet process, \cite{ferguson1973bayesian}), or considering other nonparametric models  (e.g., neural networks).  Thirdly, the proposed MTPP assumes stationarity in clinical decisions since it yields the same distribution over visitation timing and dosages given the same history. Although the stationarity is reasonable in the kidney transplantation application, our approach can be easily extended to allow non-stationary clinical decisions, by explicitly incorporating a time-varying term (e.g., $\exp(ct)$ where $c \in \mathbb{R}$ is a coefficient parameter) in  \eqref{eqn:intensity}. 
In addition, the policy gradient method is used to optimize personalized clinical decisions in this paper. Other optimization methods such as natural evolution strategies \citep{wierstra2014natural} will be explored in the second step of the proposed approach. 
Lastly, patients with chronic conditions may take multiple medicines, e.g., mycophenolate mofetil (an immunosuppressive drug) and steroids along with tacrolimus in our kidney transplantation application. 
Modeling the effects of multiple types of drugs (and their interactions with clinical observations) and learning their optimal dosage-assigning policies in the proposed optimization method will be an interesting and challenging research topic.  

\section*{Acknowledgement}
This work was supported by National Science Foundation DMS1918854 and 1940107.

\bibliography{Bib_Joint}


\end{document}